\documentclass[12pt]{article}
\usepackage{fullpage,%citesort,
  graphics,amsbsy,amssymb,amsmath,%../timestamp
}
\newcommand{\beq}{\begin{equation}}
\newcommand{\eeq}{\end{equation}}
\newcommand{\bea}{\begin{eqnarray}}
\newcommand{\eea}{\end{eqnarray}}

\newcommand{\ket}[1]{|#1\rangle}
\newcommand{\bra}[1]{\langle#1|}

\def\math{\mathsurround=0pt }
\def\leftrightarrowfill{$\math \mathord\leftarrow \mkern-6mu
 \cleaders\hbox{$\mkern-2mu \mathord- \mkern-2mu$}\hfill
 \mkern-6mu \mathord\rightarrow$}
\def\overleftrightarrow#1{\vbox{\ialign{##\crcr
     \leftrightarrowfill\crcr\noalign{\kern-1pt\nointerlineskip}
     $\hfil\displaystyle{#1}\hfil$\crcr}}}

\newcommand{\VEV}[1]{\left\langle#1\right\rangle}
\let\l=\lambda

 \def\bd{\begin{document}} \def\ed{\end{document}}
\def\ds{\documentstyle} \let\fr=\frac \let\bl=\bigl \let\br=\bigr
\let\Br=\Bigr \let\Bl=\Bigl
\let\bm=\bibitem
\let\na=\nabla
\let\pa=\partial \let\ov=\overline
\def\ft#1#2{{\textstyle{{\scriptstyle #1}\over {\scriptstyle #2}}}}
\def\fft#1#2{{#1 \over #2}}
\def\vp{\varphi}
\def\sst#1{{\scriptscriptstyle #1}}
\def\oneone{\rlap 1\mkern4mu{\rm l}}
\def\td{\tilde}
\def\wtd{\widetilde}
\def\dalemb#1#2{{\vbox{\hrule height .#2pt
        \hbox{\vrule width.#2pt height#1pt \kern#1pt
                \vrule width.#2pt}
        \hrule height.#2pt}}}
\def\square{\mathord{\dalemb{6.8}{7}\hbox{\hskip1pt}}}
\def\wtd{\widetilde}
\def\R{\rlap{\rm I}\mkern3mu{\rm R}}
\def\im{{\rm i}}
\def\tilg{\tilde{g}}
\def\tilF{\tilde{F}}
\def\tilA{\tilde{A}}
\def\varf{\varphi}
\def\tilf{\tilde{\phi}}
\def\tilh{\tilde{h}}
\def\rme{{\rm e}}
\def\ep{\epsilon}
\def\0{{(0)}}
\def\9{{(9)}}
\def\8{{(8)}}
\def\7{{(7)}}
\def\6{{(6)}}
\def\5{{(5)}}
\def\4{{(4)}}
\def\3{{(3)}}
\def\2{{(2)}}
\def\1{{(1)}}
\newcommand{\trace}{{\rm Tr}}
\newcommand{\ub}{\overline{U}}
\newcommand{\vb}{\overline{V}}
\newcommand{\uh}{\widehat{U}}
\newcommand{\vh}{\widehat{V}}
\newcommand{\ubh}{\overline{\widehat{U}}}
\newcommand{\vbh}{\overline{\widehat{V}}}
\newcommand{\lb}{\bar{\l}}
\newcommand{\Fb}{\overline{F}}
\newcommand{\Fh}{\widehat{F}}
\newcommand{\Fbh}{\overline{\widehat{F}}}
\newcommand{\Ab}{\overline{A}}
\newcommand{\Ah}{\widehat{A}}
\newcommand{\Abh}{\overline{\widehat{A}}}
\newcommand{\Gb}{\overline{G}}
\newcommand{\Gh}{\widehat{G}}
\newcommand{\Gbh}{\overline{\widehat{G}}}
\newcommand{\Pb}{\overline{P}}
\newcommand{\Ph}{\widehat{P}}
\newcommand{\Pbh}{\overline{\widehat{P}}}
\newcommand{\Qb}{\overline{Q}}
\newcommand{\Qh}{\widehat{Q}}
\newcommand{\Qbh}{\overline{\widehat{Q}}}
\newcommand{\Bb}{\overline{B}}
\newcommand{\Bh}{\widehat{B}}
\newcommand{\Bbh}{\overline{\widehat{B}}}
\newcommand{\fhns}{\hat{F}^{\rm (NS)}}
\newcommand{\fhrr}{\hat{F}^{\rm (RR)}}
\newcommand{\ahns}{\hat{A}^{\rm (NS)}}
\newcommand{\ahrr}{\hat{A}^{\rm (RR)}}
\newcommand{\hhrr}{\hat{H}^{\rm (RR)}}
\newcommand{\hchi}{\hat{\chi}}
\newcommand{\hphi}{\hat{\phi}}
\newcommand{\htau}{\hat{\tau}}
\newcommand{\cG}{{\cal G}}
\newcommand{\cGb}{\overline{{\cal G}}}
\newcommand{\cH}{{\cal H}}
\newcommand{\cP}{{\cal P}}
\newcommand{\cPb}{\overline{{\cal P}}}
\newcommand{\cQ}{{\cal Q}}
\newcommand{\cQb}{\overline{{\cal Q}}}
\newcommand{\cM}{{\cal M}}
\newcommand{\cN}{{\cal N}}
\newcommand{\cO}{{\cal O}}
\newcommand{\cD}{{\cal D}}
\newcommand{\cL}{{\cal L}}
\newcommand{\vpp}{\mbox{$\langle{\scriptstyle++}\rangle$}}
\newcommand{\vmp}{\mbox{$\langle{\scriptstyle-+}\rangle$}}
\newcommand{\vppp}{\mbox{$\langle{\scriptstyle+++}\rangle$}}
\newcommand{\vmpp}{\mbox{$\langle{\scriptstyle-++}\rangle$}}
\newcommand{\vpmp}{\mbox{$\langle{\scriptstyle+-+}\rangle$}}

\begin{document}
\begin{titlepage}
\begin{flushright}
%\phantom{UFIFT-HEP}\timestamp
\end{flushright}

\vskip1cm
\begin{center}
\begin{large}
  {\bf Heisenberg spin chain as a 
    worldsheet coordinate\\ for lightcone quantized string
 %%%  Overlaps in Free Fermion and Spin  Systems
  }
\end{large}

\vskip 2cm
{
Charles B. Thorn\footnote{E-mail  address: {\tt thorn@phys.ufl.edu}}
}
\vskip0.20cm
{\it Institute for Fundamental Theory\\
Department of Physics, 
University of Florida
Gainesville FL 32611}

%(\today)

\vskip 1.0cm
\end{center}

\begin{abstract}\noindent
  Although the energy spectrum of the Heisenberg spin chain on a circle
  defined by 
  \bea
H&=&\frac{1}{4}\sum_{k=1}^M(\sigma_k^x\sigma_{k+1}^x+\sigma_k^y\sigma_{k+1}^y
+\Delta\sigma_k^z\sigma_{k+1}^z)
\eea
 is well known for any  fixed $M$,
  the boundary conditions vary according to whether $M\in 4\mathbb{N}+r$,
  where $r=-1,0,1,2$, and also according to the parity of the number of
  overturned spins in the state,
  In string theory all these cases must be allowed
  because interactions involve a string with $M$ spins breaking into
  strings with $M_1<M$ and $M-M_1$ spins (or vice versa).
  We organize the energy spectrum and degeneracies of $H$
  in the case $\Delta=0$ where the system is equivalent to a
  system of free fermions. In spite of the multiplicity of special cases,
  in the limit $M\to\infty$ the
  spectrum is that of a free compactified worldsheet field. Such a field
  can be interpreted as a compact transverse string coordinate $x(\sigma)
  \equiv x(\sigma)+R_0$. We construct the bosonization formulas explicitly
  in all separate cases, and for each sector give the Virasoro conformal
  generators in both  fermionic and bosonic formulations.
  Furthermore
  from calculations in the literature for selected classes of excited states,
  there is strong evidence that the only change for $\Delta\neq0$ is
  a change in the
  compactification radius $R_0\to R_\Delta$.
  As $\Delta\to-1$ this radius goes to infinity,
  giving a concrete example of  noncompact space emerging
  from a discrete dynamical system.
  Finally we apply our work to construct the three string vertex
  implied by a string
  whose bosonic coordinates emerge from this mechanism.
\end{abstract}
\vfill

\end{titlepage}
\newpage
\section{Introduction}
This article is a technical follow up to \cite{spacebits}, which proposed
a string bit model \cite{gilest,thornsakh}
for superstring \cite{gso} in which space is not fundamental, but is
rather an emergent feature of the string bit dynamics. Briefly, a
string bit\footnote{Since it is a quantum entity with a finite number of
states, it might be thought of as a $q$-byte.}
is described by annihilation (and creation) operators, which are $N\times N$
matrices $(\phi^A_{[a_1,\ldots a_n]})_\alpha^{\ \beta}$ 
($(\phi^A_{[a_1,\ldots a_n]})_\alpha^{\ \beta\dagger}$), 
where $\alpha,\beta=1,\ldots,N$ are the 
matrix indices, $a_1,\cdots, a_n$ are spinor indices  each taking
values $1,\dots,s$, with $n$ ranging from $0$ to $s$.
The $\phi$ are completely antisymmetric in these spinor indices.
The index $A$ labels
additional string bit internal states in some models.

In these models the emergence of space starts 
with the emergence of a worldsheet in 't Hooft's limit $N\to\infty$
\cite{thooftlargen}.
In this
limit the bits form into noninteracting closed chains. The lowest energy
chains have an infinite number of bits in certain models 
\cite{sunthorn,chensun,songge}: then low
energy  chains
look like strings moving in one space dimension. This one spatial
dimension emerges as the conjugate to bit number $M$, which for
$M\to\infty$ can be treated as a positive continuous variable and
interpreted as the lightcone \cite{goddardgrt} longitudinal
momentum $P^+/m$, where $m$ is the $P^+$ carried by each string bit.
This emergent coordinate is therefore appropriately named $x^-$.

However, instead of transverse spatial dimensions,
fluctuations in the string bit internal states,
labeled by the $a$ and $A$ indices, become fermionic or
compact bosonic worldsheet fields.
For example, the spinor indices on the string bit operator are promoted,
on large bit number chains 
\cite{bergmantsubit} to $s$ worldsheet Fermi fields of
the Green-Schwarz type \cite{bardakcih}. If the label
$A$ is absent, the  quantum overlap
between a two string state and a one string state
is consistent with Lorentz covariance ($SO(1,1)$) only if $s=24$
\cite{thornprotobits}. I call this string in 1 space 1 time dimensions the
protostring. It has an interesting degree of freedom count: 16 of the
24 Grassmann dimensions can be bosonized to 8 compactified transverse
boson dimensions leaving 8 Grassmann dimensions. This is
the worldsheet field content of the superstring.
Of course the familiar super Poincar\'e covariant superstring
requires 8 noncompact transverse dimensions
in addition to an operator applied to the overlap at the
string separation point.

One way for noncompact dimensions to arise from fluctuations of a finite valued
index is to appeal to the one dimensional Heisenberg spin chain.
I noted in \cite{spacebits}
that the degrees of freedom labeled by $A$
could be engineered so that the 't Hooft $N\to\infty$ limit
produces an additional
worldsheet subsystem describing a Heisenberg
spin chain with Hamiltonian
\bea
H&=&\frac{1}{4}\sum_{k=1}^M(\sigma_k^x\sigma_{k+1}^x+\sigma_k^y\sigma_{k+1}^y
+\Delta\sigma_k^z\sigma_{k+1}^z)
\label{heiham}
\eea
where $M$ is the number of bits in the chain,
In fact one can design the string bit Hamiltonian to
produce several independent
Heisenberg spin chains. Although the notation suggests that the $\sigma$'s
refer to physical spin, we shall
refer to the eigenvalues of $\sigma^z$ as ``charge''.
This Hamiltonian has been very well studied.
Eigenstates of 
$H$ (probably all of them) can be found via the
Bethe ansatz \cite{bethe}, which exploits the fact that the total charge 
$Q=\sum_{k=1}^M\sigma_k^z$ commutes with $H$, so that an eigenstate of $H$
may be assumed to have a fixed charge given by $Q=M-2q$ where
$q$  is the number of overturned spins compared
to a reference state with all spins up. Yang and Yang \cite{yangyang}
solved the Bethe ansatz equations for the ground state in each charge sector,
in the limit $M\to\infty$. For $\Delta$ in the range $-1<\Delta<1$,
the energy excitations were shown to be of order $O(M^{-1})$ corresponding
to a nontrivial continuum``stringy'' limit.
Earlier Lieb, Schultz, and Mattis had solved the $\Delta=0$ case exactly
\cite{liebsm}, based on the fact that 
the Heisenberg spin system is then equivalent to a system of free fermions. 
A discussion of the eigenstates (with $E=O(M^{-1})$ 
relevant to the interpretation of the system as a compactified worldsheet
coordinate is given in \cite{gilesmt,spacebits}.
The compactification radius $R$ depends on $\Delta$:
\bea
R^2_\Delta=\frac{\pi\alpha^\prime}{\mu},\qquad \Delta=-\cos\mu .
\eea
Here the Regge slope parameter $\alpha^\prime=1/(2\pi T_0)$,  where
$T_0$ is the rest tension of the string determined by the excitation
energy spectrum.
This interpretation is solid when $\Delta=0$, but for $\Delta\neq0$ it relies
on   the Bethe ansatz and the calculations in \cite{yangyang} for some
selected classes of excitation.
These included the lowest states with any fixed $Q$ and ${\hat P}/\pi$
and spin wave excitations relative to these lowest states. Here ${\hat P}
=P-\pi q$ with $P$ the total lattice momentum {\it modulo} $O(M^{-1})$
as $M\to\infty$.
For $\Delta\neq0$
it is logically possible that the Bethe ansatz misses some eigenstates,
which could also be missed by the compactified coordinate interpretation.

The equivalence of the Heisenberg spin system to a compactified bosonic
worldsheet field when $-1<\Delta<1$ provides a shortcut to the calculation of
string scattering amplitudes: the string technology for such worldsheet
systems is very well developed. However there remains some doubt that the
bosonization formalism really captures all aspects of the Heisenberg system at
large $M$. The purpose of this article is to flesh out the connection between
the original spin system and its bosonized counterpart. It may be that resort
to numerical methods may be needed. So we lay out in painstaking detail the
Fermi Bose connection at the free fermion point $\Delta=0$. This is to test
the two descriptions of the string interactions.

This work is quite tedious because the worldsheet fermionic fields
of the system have different boundary conditions for different values
of the charge $Q$. For example, assuming that $M$ is a multiple of 4,
the fermionic fields are antiperiodic when $Q$ is also a multiple of 4,
but periodic when $Q$ is an odd multiple of 2. This
assignment is reversed if $M$ is an odd multiple of 2. When $M$ is odd
the situation is even more involved. In string theory the interactions
change the number of bits in a string by breaking one string into two or joining two strings to form 1. Since bit number is conserved, the three strings in the process cannot have the same number of bits. One must allow strings with any number of bits.
Thus one must do a separate calculation
for 8 different sectors for each of the three chains participating
in the vertex. We set up the vertex construction in a way to cover any of
these cases, 

In spite of all the different separate cases, in the limit $M\to\infty$
the final result for the spectrum of the Heisenberg Hamiltonian at the
free fermion point is simply described in the language of bosonization:
the energy spectrum is identical to that of a bosonic worldsheet
field compactified on a
circle of a specific radius $R_0$ \cite{gilesmt}.
Thus the contribution to the string vertex at $\Delta=0$ can, in principle,
be evaluated
in two ways: either in the free fermion language, or the free boson
language. Whether these two methods give identical results is not
clear. Unfortunately we are so far unable
to obtain the $M\to\infty$
limit for the fermionic version for reasons we shall explain.
But since bosonization is only valid  in this limit, a direct comparison 
cannot yet be made.

This bosonized language makes the
$\Delta\neq0$ case tractable, For the energy spectrum and degeneracies
for $M$ large 
then includes that of
a boson compactified on a circle of different radius $R_\Delta$. 
Interestingly there is a limit ($\Delta\to-1$)
in which $R\to\infty$, yielding an
uncompactified coordinate, which can be identified as one of the transverse
coordinates of a string moving  in flat space.
For the superstring we would need 8 such Heisenberg systems along with
the 8 Fermi
Grassmann dimensions that
emerge from fluctuations in the spinor indices.
While
the fermionic Hamiltonian has a quartic interaction term
proportional to $\Delta$, the bosonic
form of the Hamiltonian is quadratic for all $\Delta$,
and we can calculate the overlap using the
methods of Mandelstam \cite{mandelstamlc}, Indeed, for the decompactification
limit Mandelstam's three string vertex is the desired answer.

In Section 2 we set up overlap calculations of the sort required in
string theory for general free fermion systems.
In sections 3 and 4 we do the painstaking, case by case, work
of enumerating the (already known)
energy eigenstates and eigenvalues for the
Heisenberg spin system at $\Delta=0$. We summarize all the separate cases
for ground states in a table. Then we describe the complete spectrum and
degeneracies in the large $M$ limit, using partition functions.
In Section 4 we further
characterize the large $M$ limit by constructing the Virasoro generators
and the operator bosonization formulas. Finally in Section 5 we apply the
results of Section 2 to construct the overlap at $\Delta=0$ for finite $M$,
in terms of an $M\times M$ matrix. A submatrix of this matrix must be inverted
to obtain the overlap. This can of course be done numerically for finite $M$.
We give the large $M$ limit of this
matrix, and explain why Goldstone's method (at least in its original form)
for inverting the matrix,
which succeeds for the bosonic string and Green-Schwarz superstring overlap
\cite{brinkgs},
falls short. Nonetheless, we can see from the limit that left and right moving
spin waves do decouple from each other, as they should.  
This feature is shared by
the overlap in bosonic language. The latter can be extended
to $\Delta\neq0$, allowing Mandelstam's methods to be applied
to the general case.

\section{Overlaps in Free Fermion Systems}
We are interested in systems that undergo a sudden change of Hamiltonian at
various times, as occurs in the worldsheet description of string theory.
But here we concentrate on just a single sudden change in a system of fermions
described by anticommuting energy annihilation and creation
operators. So for the
first (second) Hamiltonian we introduce annihilation operators $B_k$ ($b_k$)
and creation operators $B^\dagger_k$  ($b^\dagger_k$):
\bea
{}[H_1, B_k]&=& -\Omega_k B_k,\qquad \{B_k,B^\dagger_l\}=\delta_{kl}\\
{}[H_2, b_k]&=& -\omega_k b_k,\qquad \{b_k,b^\dagger_l\}=\delta_{kl}
\eea
We shall not assume that the energies $\omega, \Omega$ are positive.
So, as in the Dirac system, the ground states have negative energy
levels occupied. We shall assume there is a common reference eigenstate
$\ket{0}$ annihilated by both $B_k$ and $b_k$, as well as by $H_1$ and $H_2$.
(This is a feature of the Heisenberg spin system at the free fermion point,
and it's also a feature of the Dirac free fermion system.)
Then any eigenstate of $H_1$ or $H_2$ can be obtained from the coherent states
\bea
\ket{\gamma}&=&e^{\sum_s\gamma_s B^\dagger_s}\ket{0}\\
\ket{\alpha}&=&e^{\sum_r\alpha_r b^\dagger_r}\ket{0}
\eea
where $\alpha,\gamma$ are Grassmann numbers. Let $G$ be the set of $s$
for which $\Omega_s<0$ and $g$ be the set of $r$ for which $\omega_r<0$. Then
the ground states can be expressed as
\bea
\ket{G}&=& \prod_{s\in G}\frac{d}{d\gamma_s}\ket{\gamma}|_{\gamma=0}
=\prod_{s\in G}B_s^\dagger\ket{0}\\
\ket{g}&=& \prod_{r\in g}\frac{d}{d\alpha_r}\ket{\alpha}|_{\alpha=0}
=\prod_{r\in g}b_r^\dagger\ket{0}
\eea
The ordering of the operators applied to a ket
is set by convention to be increasing index from left to right. This ordering
is reversed in $\bra{G},\bra{g}$. The bra coherent states are written
\bea
\bra{\gamma}&=&\bra{0}e^{\sum_s {\gamma}_sB_s}\\
\bra{\alpha}&=&\bra{0}e^{\sum_r{\alpha}_r b_r}
\eea
so in obtaining $\bra{G}$ and $\bra{g}$ from the coherent states
one just reverses the ordering of the derivatives relative to the ket
construction. The overlaps of interest to us are between states obtained
by applying monomials of  positive energy creation operators and
negative energy annihilation operators  to $\ket{G}$ or $\ket{g}$.
In terms of coherent states
\bea
\ket{\gamma,{\hat\gamma}}=e^{\sum_{s\notin G}\gamma_s B^\dagger_s }
e^{\sum_{s\in G}{\hat\gamma}_s B_s }\ket{G}\\
\ket{\alpha,{\hat\alpha}}=e^{\sum_{r\notin g}\alpha_s
  b^\dagger_r }
e^{\sum_{r\in g}{\hat\alpha}_r b_r }\ket{g}
\eea
The annihilation operators create holes in the negative energy sea.

We are interested in matrix elements
\bea
\VEV{g\left|e^{\sum_{r\in g}{\hat\alpha}_r b^\dagger_r }e^{\sum_{r\notin g}\alpha_r
  b_r }e^{\sum_{s\notin G}\gamma_s B^\dagger_s }
e^{\sum_{s\in G}{\hat\gamma}_s B_s }\right|G}
\nonumber
\eea
which can be computed by taking derivatives of
\bea
&&\hskip-.5in\VEV{0\left|e^{\sum_{r\in g}\alpha_r b_r}
      e^{\sum_{r\in g}{\hat\alpha}_r b^\dagger_r }
    e^{\sum_{r\notin g}\alpha_r
  b_r }e^{\sum_{s\notin G}\gamma_s B^\dagger_s }
e^{\sum_{s\in G}{\hat\gamma}_s B_s }e^{\sum_{s\in G}\gamma_s B^\dagger_s }
\right|0}\nonumber\\
&=& e^{\sum_{r\in g}{\hat\alpha}_r\alpha_r+\sum_{s\in G}\gamma_s
    {\hat\gamma}_s }\VEV{0\left|
e^{\sum_{r}\alpha_r b_r}
    e^{\sum_{s}\gamma_s B^\dagger_s }
  \right|0}\nonumber\\
&=& e^{\sum_{r\in g}{\hat\alpha}_r\alpha_r+\sum_{s\in G}\gamma_s
    {\hat\gamma}_s }\VEV{0\left|
    e^{\sum_{s}\gamma_s\alpha_r \{b_r.B_s^\dagger\}}
  \right|0}= e^{\sum_{r\in g}{\hat\alpha}_r\alpha_r+\sum_{s\in G}\gamma_s
    {\hat\gamma}_s +\sum_{s}\gamma_s\alpha_r D_{sr}}
\label{coherent}
\eea
In these fermion models $\{b_r.B_s^\dagger\}=D_{sr}$ is a numerical matrix.
This matrix also gives the relation between the two bases:
\bea
B_s^\dagger=D_{sr}b_r^\dagger,\qquad B_s=D^*_{sr}b_r=b_rD^\dagger_{rs}
\eea
The anticommutation relations imply that $D$ is a unitary matrix. If we think
of $B,b$ as row vectors and $B^\dagger, b^\dagger$ as column vectors, one
can safely suppress indices in some equations, e.g.
\bea
B^\dagger=Db^\dagger,\qquad B=bD^\dagger,\qquad
b^\dagger=D^\dagger B^\dagger,\qquad b=BD.
\eea
To construct the ket $\ket{G}$ we simply apply a product of derivatives w.r.t.
$\gamma_s$ with $s\in G$ after which we set these $\gamma$ to zero.
\bea
&&\hskip-.5in\VEV{0\left|e^{\sum_{r\in g}\alpha_r b_r}
      e^{\sum_{r\in g}{\hat\alpha}_r b^\dagger_r }
    e^{\sum_{r\notin g}\alpha_r
  b_r }e^{\sum_{s\notin G}\gamma_s B^\dagger_s }
e^{\sum_{s\in G}{\hat\gamma}_s B_s }\right|G}\nonumber\\
&=&e^{\sum_{r\in g}{\hat\alpha}_r\alpha_r+\sum_{s\notin G}
  \gamma_s\alpha_r D_{sr}}
\prod_{s\in G}\left({\hat\gamma}_s+\sum_r\alpha_r D_{sr}\right)
\nonumber
\eea
to construct the bra $\bra{g}$ we again apply a product of derivatives, this
time w.r.t. $\alpha_r$, but the result is more complex because the product
factors also contain $\alpha_r$ with $r\in g$.
\bea
&&\hskip-.5in\VEV{g\left|
      e^{\sum_{r\in g}{\hat\alpha}_r b^\dagger_r }
    e^{\sum_{r\notin g}\alpha_r
  b_r }e^{\sum_{s\notin G}\gamma_s B^\dagger_s }
e^{\sum_{s\in G}{\hat\gamma}_s B_s }\right|G}\nonumber\\
&=&e^{\sum_{s\notin G, r\notin g}
  \gamma_s\alpha_r D_{sr}}
\prod_{r\in g}\left(\frac{d}{d\alpha_r}-{\hat\alpha}_r-\sum_{s\notin G}\gamma_s D_{sr}\right)\prod_{s\in G}\left({\hat\gamma}_s+\sum_r\alpha_r D_{sr}\right)
\nonumber
\eea
where after the derivatives are performed, $\alpha_r$ for $r\in g$
is set to zero. This looks daunting, However, because we are dealing with free fermions, we only need the first two terms in the expansion of the product of
derivatives about $x_r\equiv -{\hat\alpha}_r-\sum_{s\notin G}
\gamma_s D_{sr}=0$: 
\bea
\prod_{r\in g}\left(\frac{d}{d\alpha_r}+x_r\right)&=&
\prod_{r\in g}\left(\frac{d}{d\alpha_r}\right)+\sum_t x_t(-)^{n-t}
\prod_{r\in g r\neq t}\left(\frac{d}{d\alpha_r}\right)+O(x^2)
\eea
The reason is that in a diagrammatic expansion in a free fermion system the
only connected diagram is the two point function, and the sum of all diagrams
is just the exponential of the connected diagrams. The first term in the
expansion about $x=0$ is the zero point function and the second  nonzero
term is the two point function.

Assuming that the sets $G,g$ are the same size,
the first term gives
\bea
\prod_{r\in g}\left(\frac{d}{d\alpha_r}\right)\prod_{s\in G}
\left({\hat\gamma}_s+\sum_r\alpha_r D_{sr}\right)&=&
\sum_P(-)^P\prod_{s\in G, P_s\in g}D_{s,P_s}=\det D^{Gg}
\eea
where $D^{Gg}$ is the matrix formed by restricting the matrix $D_{sr}$
to $r\in g$ and $s\in G$.

The second term gives
\bea
&&\hskip-.5in \sum_{t\in g} x_t(-)^{n-t}
\prod_{r\in g,r\neq t}\left(\frac{d}{d\alpha_r}\right)
\prod_{s\in G}
\left({\hat\gamma}_s+\sum_r\alpha_r D_{sr}\right)\nonumber\\
&=&
\sum_{t\in g} x_t(-)^{n-t}\sum_{u\in G}(-)^{u+n}
\left({\hat\gamma}_u+\sum_{r\notin g}
  \alpha_{r} D_{ur}\right)
\prod_{r\in g,r\neq t}\left(\frac{d}{d\alpha_r}\right)
\prod_{s\in G, s\neq u}
\left(\sum_{r\in g}\alpha_r D_{sr}\right)
\nonumber\\
&=&
\sum_{t\in g,u\in G} (-)^{u-t}\left(-{\hat\alpha}_t-\sum_{s\notin G}
\gamma_s D_{st}\right)\left({\hat\gamma}_u+
  \sum_{r\notin g}
  \alpha_{r} D_{ur}\right)\det m^{Gg}_{ut}
\eea
where $m^{Gg}_{ut}$ is the minor matrix formed from $D^{Gg}$ by deleting
the row and column containing $D_{ut}$. 
By standard construction, the matrix $A_{tu}=(-)^{u-t}
\det m^{Gg}_{ut}/\det D^{Gg}$
is the inverse of the matrix $D^{Gg}$. Thus we can write
\bea
&&\hskip-.3in\VEV{g\left|
      e^{\sum_{r\in g}{\hat\alpha}_r b^\dagger_r }
    e^{\sum_{r\notin g}\alpha_r
  b_r }e^{\sum_{s\notin G}\gamma_s B^\dagger_s }
e^{\sum_{s\in G}{\hat\gamma}_s B_s }\right|G}\nonumber\\
&=&\det D^{Gg} e^{\sum_{s\notin G, r\notin g}
  \gamma_s\alpha_r D_{sr}}\left[1\phantom{\sum_{r\notin g}}\right.\nonumber\\
  &&\left. +\sum_{t\in g,u\in G}
  \left({\hat\gamma}_u+
  \sum_{r\notin g}
  \alpha_{r} D_{ur}\right)(D^{Gg})^{-1}_{tu}
\left({\hat\alpha}_t+\sum_{s\notin G}
\gamma_s D_{st}\right)+\cdots
\right]\nonumber\\
&=&\det D^{Gg}\exp\left[\sum_{s\notin G, r\notin g}
  \gamma_s\alpha_r D_{sr}\phantom{\sum_{r\notin g}}\right.\nonumber\\
&&\left.+\sum_{t\in g,u\in G}
  \left({\hat\gamma}_u+
  \sum_{r\notin g}
  \alpha_{r} D_{ur}\right)(D^{Gg})^{-1}_{tu}
\left({\hat\alpha}_t+\sum_{s\notin G}
\gamma_s D_{st}\right)
\right]
\label{ffoverlap}
\eea
The dots within square brackets in the second line signify terms quartic and
higher in Grassmann variables, and writing the contents of those square
brackets as an exponential is valid because this is a system of free fermions.
This formula gives all of the overlaps of interest in terms of the matrix
$D_{rs}$ \footnote{This formula (\ref{ffoverlap}) can be obtained,
simply and transparently,  by integrating (\ref{coherent}) over $\gamma_s$ with $s\in G$ and over
$\alpha_r$ with $r\in g$. I thank a referee for reminding me of the
equivalence between Grassmann integration and differentiation in
this context.}.
%%%%
\section{Heisenberg Spin System}
%%%%
\subsection{Hamiltonian in fermionic language}
We write the Hamiltonian for the Heisenberg one dimensional spin system
as a nominally antiferromagnetic system with $M$  spins
\bea
H&=&\frac{1}{4}\sum_{k=1}^M(\sigma_k^x\sigma_{k+1}^x+\sigma_k^y\sigma_{k+1}^y
+\Delta\sigma_k^z\sigma_{k+1}^z)
\eea
and impose periodic boundary conditions $\sigma_{M+1}\equiv\sigma_1$.
The $\sigma_k$'s are Pauli spin matrices independent on each site. They satisfy
the commutator algebra
\bea
{}[\sigma^a_k,\sigma^b_l]=0,\quad l\neq k;\qquad \{\sigma^a_k,\sigma_k^b\}
=2\delta_{ab}.
\eea
While the notation suggests the spin interpretation of the 
Pauli matrices, they could easily represent an internal symmetry like
isospin. That's how we regard them in the string bit models.
An important symmetry of the Heisenberg model is
generated by the charge $Q=\sum_k\sigma_k^z$, which commutes
with $H$ because rotational symmetry about the $z$ axis is unbroken by
$\Delta\neq0$. This symmetry underlies the success of the Bethe ansatz.

The Jordan-Wigner trick defines anticommuting variables $S_k^a$, $a=x,y$
by
\bea
S_k^{x,y}=\sigma_k^{x,y}\prod_{l=1}^{k-1}\sigma_l^z,
\eea
so that
\bea
\{S_k^a,S_l^b\}=2\delta_{kl}\delta_{ab},\qquad a,b\neq z.
\eea
Using $\sigma_k^a\sigma_k^b=\delta_{ab}+i\epsilon_{abc}\sigma_k^c$ one can
express $H$ entirely in terms of the $S$'s. 
Then the $\Delta$ term is
quartic in the $S$'s, so the free fermion case corresponds to $\Delta=0$
which we assume in the rest of this section and all of the next section.
\subsection{$\Delta=0$: Free fermion case}
The Hamiltonian then becomes
\bea
H&=&\frac{i}{4}\sum_{k=1}^{M-1}(S^x_kS^y_{k+1}-S^y_kS^x_{k+1})-\frac{i}{4}
(S_M^x S_1^y-S_M^y S_1^x)\Omega
\eea
where $\Omega=\prod_{k=1}^M\sigma^z_k$. We find it convenient to
use raising and lowering operators $S_k=(S_k^x+iS_k^y)/{2}$
and $S^\dagger_k$, satisfying $\{S_k,S_l\}=0$ and
$\{S_k,S^\dagger_l\}=\delta_{kl}$. The $H$ reads
\bea
H&=&\frac{1}{2}\sum_{k=1}^{M-1}(S^\dagger_kS_{k+1}+S^\dagger_{k+1}S_{k})
-\frac{1}{2}(S_M^\dagger S_1+S_1^\dagger S_M)\Omega.
\eea
Clearly the boundary conditions depend on the value  of $\Omega=\pm1$.
The energy raising and  lowering operators are just Fourier transforms
of the $S$'s.
\bea
S_k=\frac{1}{\sqrt{M}}\sum_r e^{2\pi ikr/M}B_r,\qquad B_r=
\frac{1}{\sqrt{M}}\sum_k e^{-2\pi ikr/M}S_k
    \eea
    Here $r=n+\alpha$, with $n$ an integer in $0\leq n\leq M-1$.
    The fractional part of $r$ determines the boundary conditions on $S_k$:
    \bea
    S_{M+1}=e^{2\pi i\alpha}S_1
    \eea
    Thus $\alpha=0$ for $\Omega=-1$ and $\alpha=1/2$ for $\Omega=+1$.
    In other words
    $r$ ranges through integers (half odd integers) if $\Omega=-1$
    ($\Omega=+1$). Plugging the expansions for $S,S^\dagger$ into $H$
    gives
    \bea
    H=\sum_r\cos\frac{2\pi r}{M}B^\dagger_r B_r,
    \qquad \{B_r.B^\dagger_s\}=\delta_{rs}
    \eea
    Thus $B^\dagger_r$ creates an amount of energy $\cos(2\pi r/M)$, which is
    negative for $r$ in the range $M/4<r<3M/4$. In some cases $r$
    can be exactly $M/4$ or $3M/4$, in which case the corresponding
    $B^\dagger$ creates zero
    energy. In constructing the ground state, all negative energy levels
    usually must be occupied\footnote{Because for fixed $\Omega$ excited
      states always involve at least two levels, it is possible that the lowest
      energy state involves an unoccupied negative energy level or an occupied
    positive energy level, This happens in the spin system when $M$ is odd.},
    but the existence of zero energy modes means the
    ground level is degenerate, and the selection of ground state is ambiguous,
    depending on how many of the zero energy levels are filled.
    The nonzero energy levels with $r\neq M/2$ are doubly degenerate
    due to the symmetry under $r\to M-r$.
\subsection{Dependence of ground energy on the nature of $M$}
The Heisenberg spin chain is antiferromagnetic in character for all values
$-1<\Delta<+1$. This means that the ground state tends to minimize
$|Q|$. Starting with the state with all spins up ($Q=M$), by overturning
$q$ of the spins the charge is lowered to $Q=M-2q$.
When $M$ is even, the ground state has $Q=0$ because
one can overturn exactly half of the spins in the state $\ket{0}$.
However when $M$ is odd, this possibility is frustrated and the ground
state has $Q=\pm1$. But in each of these cases there are two subcases.

For $M$ even they are $M=$ a multiple of 4, with an
even number of negative energy states filled with half odd integer modes;
and  $M= 2$ + a multiple of 4, with an odd number of negative energy states
filled with integer modes. The table shows that when the mode numbers in these
two cases are expressed in terms of $M$, the patterns in the two cases
are identical and in particular the energies are identical functions of $M$.
The same story holds for the lowest energy states with opposite value of
$\Omega$ to the ground state. The degeneracy between $Q=\Sigma=\pm2$ is
due to time reversal invariance. It can also be seen by applying the
operator $B^\dagger_{M/4}B^\dagger_{3M/4}$, which creates 0 energy and destroys 4 units of charge, to the $Q=+2$ ground state.

\begin{table}[ht]
\begin{center}
\begin{tabular}{||c||c|c|c|c|c|c||}
\hline\hline
M & G & q& $Q$&$\frac{{\hat P}}{\pi}$& $\Omega$& $E_G$\\
\hline \hline
  $M\in 4\mathbb{N}$& $\frac{M+2}{4},\cdots,\frac{3M-2}{4}
                      \in
                      \frac{1}{2}+\mathbb{Z}$& $\frac{M}{2}$, even&0&0&+1
  &$-{\csc\frac{\pi}{M}}$\\
  \hline\hline
  & $\frac{M+4}{4},\cdots,\frac{3M-4}{4}
    \in
               \mathbb{Z}$& $\frac{M-2}{2}$, odd&+2&0&-1&$-\cot\frac{\pi}{M}$\\
  \hline
 & $\frac{M}{4},\cdots,\frac{3M}{4}
    \in
            \mathbb{Z}$& $\frac{M+2}{2}$, odd&-2&0&-1&$-\cot\frac{\pi}{M}$\\
  \hline\hline
  $M\in 4\mathbb{N}+2$ & $\frac{M+2}{4},\cdots,
                         \frac{3M-2}{4}\in\mathbb{Z}$
      &$\frac{M}{2}$,                                                                                 odd&0&0&-1&$-{\csc\frac{\pi}{M}}$\\
  \hline\hline
   & $\frac{M+4}{4},\cdots,\frac{3M-4}{4}
    \in
     \frac{1}{2} +  \mathbb{Z}$& $\frac{M-2}{2}$,
                                 even&+2&0&+1&$-\cot\frac{\pi}{M}$\\
  \hline
 & $\frac{M}{4},\cdots,\frac{3M}{4}
    \in
   \frac{1}{2}+   \mathbb{Z}$& $\frac{M+2}{2}$,
                               even&-2&0&+1&$-\cot\frac{\pi}{M}$\\
  \hline\hline
  $M\in 4\mathbb{N}+1$ & $\frac{M+1}{4},\cdots,\frac{3M-5}{4}
                        \in
                      \frac{1}{2}+\mathbb{Z}$  &$\frac{M-1}{2}$,
                      even&+1&$-\frac{q}{M}$&+1&$-\cot\frac{\pi}{M}\cos\frac{\pi}{2M}$\\
  \hline
  & $\frac{M+5}{4},\cdots,\frac{3M-1}{4}\in
                      \frac{1}{2}+\mathbb{Z}$  &$\frac{M-1}{2}$,
                                                 even&+1&$\frac{q}{M}$&+1
  &$-\cot\frac{\pi}{M}\cos\frac{\pi}{2M}$\\
  \hline\hline
  & $\frac{M+3}{4},\cdots,\frac{3M+1}{4}
            \in\mathbb{Z}$  &$\frac{M+1}{2}$,
   odd&-1&$\frac{q}{M}$&-1&$-\cot\frac{\pi}{M}\cos\frac{\pi}{2M}$\\
  \hline
  & $\frac{M-1}{4},\cdots,
    \frac{3M-3}{4}\in\mathbb{Z}$  &$\frac{M+1}{2}$,
                                              odd&-1&$-\frac{q}{M}$&-1
                                &$-\cot\frac{\pi}{M}\cos\frac{\pi}{2M}$\\
\hline\hline
  $M\in 4\mathbb{N}-1$ & $\frac{M+1}{4}
                        ,\cdots,\frac{3M-5}{4}\in
                         \mathbb{Z}$  &$\frac{M-1}{2}$, odd&+1
              &$-\frac{q}{M}$&-1
                              &$-\cot\frac{\pi}{M}\cos\frac{\pi}{2M}$\\
  \hline
  & $\frac{M+5}{4},\cdots,\frac{3M-1}{4}\in
                       \mathbb{Z}$  &$\frac{M-1}{2}$,
         odd&+1&$\frac{q}{M}$&-1&$-\cot\frac{\pi}{M}\cos\frac{\pi}{2M}$\\
  \hline\hline
  & $\frac{M+3}{4},\cdots,\frac{3M+1}{4}\in\frac{1}{2}
    +\mathbb{Z}$  &$\frac{M+1}{2}$,
                    even&-1&$\frac{q}{M}$&+1&
                              $-\cot\frac{\pi}{M}\cos\frac{\pi}{2M}$\\
  \hline
  & $\frac{M-1}{4},\cdots,\frac{3M-3}{4}
    \in\frac{1}{2}+\mathbb{Z}$  &$\frac{M+1}{2}$,
                                              even&-1&$-\frac{q}{M}$&+1
                             &$-\cot\frac{\pi}{M}\cos\frac{\pi}{2M}$\\
\hline\hline
\end{tabular}
\caption{
  Momentum modes contributing to the ground state(s) in both $\Omega=\pm1 $
  sectors for fixed $M$, together with values of $q,Q,{\hat P/\pi}$, and energy
  in each case.} 
\label{groundmomenta}
\end{center}
\end{table}

When $M$ is odd we again have two cases for which $M$ is $\pm1$ plus a
multiple of 4. The $\Omega$'s are opposites, but energy patterns are identical
when expressed in terms  of $M$. For a fixed $M$ there are 4 degenerate states,
two with $Q=+1$ and two with $Q=-1$. Again the degeneracy of the
$Q=\pm1$ states is a consequence of time reversal, but since they have opposite
$\Omega$'s there is no simple operator linking them as in the even case.
But the two states with the same charge are linked by simple operators:
$B^\dagger_{(3M-1)/4}B_{(M+1)/4}$ takes line 7 to line 8
(or line 11 to line 12) in the table,
while $B^\dagger_{(M-1)/4}B_{(3M+1)/4}$
takes line 9 to line 10 (or line 13 to line 14). Both of these operators
create 0 energy, because they create and destroy two states of the same energy:
\bea
\cos\frac{2\pi(3M-1)}{4M}=-\sin\frac{\pi}{2M}=\cos\frac{2\pi(M+1)}{4M}.
\eea
\subsection{Energy eigenstates for $\Delta=0$,}
The description of the complete set of energy eigenstates is
complicated by the need to switch mode numbers from half odd integers
to integers when
$\Omega$ changes from $+1$ to $-1$. The ground states delineated in
the previous subsection can form convenient starting eigenstates. For fixed
$M$ one must pick a ground state in each of the two $\Omega$ sectors. Then a
complete set of states is generated by applying even monomials of
eigenoperators, moded appropriately, to each of the choices. When there is a
degeneracy of ground states in a given sector
one can pick one of them at will. For
definiteness, we choose lines 1,2,4,5,7,9,11,and 13 as our ground states.
Then the even monomials will consist of half odd integer moded
eigenoperators for cases 1,5,7,13 and integer moded eigenoperators for
cases 2,4,9,11.

The momentum carried by a mode $r$ is $p_r=2\pi r/M$.
This relation is used to calculate the entries in the table under
${\hat P}/\pi$. ${\hat P}\equiv P-\pi q$, where P is the total momentum
of the system. It commutes with the Hamiltonian because there is a discrete
lattice translational invariance with periodic boundary conditions.
At the level of individual modes
the definition of ${\hat P}$
corresponds to ${\hat p}_r\equiv p_r-\pi$. The ${\hat p}_r$ distribution is
centered around $0$ instead of around $\pi$.
The eigenvalue of ${\hat P}$ distinguishes
the degenerate ground states with equal energy and charge.

One can construct any excited state by applying combinations of
bilinears of creation operators to the ground state in a given sector.
For this purpose, creation operators are $B^\dagger_r$ for $r\notin G$ and
$B_s$ for $s\in G$. then the possible bilinears  are:
\bea
B^\dagger_rB^\dagger_s,\qquad B^\dagger_rB_t,\qquad B_tB_u,\qquad
r,s\notin G;\quad t,u\in G
\eea
If two or more bilinears are applied, the state will be nonzero only
if all indices are different.  Furthermore two nonzero states are proportional
if the sets of indices are permutations of each other. The listed bilinears
change $Q$ by -4, 0, +4 respectively; ${\hat P}$ by $2\pi[-1+(r+s)/M]$,
$2\pi(r-t)/M$, $-2\pi[-1+(t+u)/M]$ respectively; and
$E$ by $\cos(2\pi r/M)+\cos(2\pi s/M)$, $\cos(2\pi r/M)-\cos(2\pi t/M)$,
$-\cos(2\pi t/M)-\cos(2\pi u/M)$ respectively.

To get stringy physics, we should take $M\to\infty$, and concentrate on
low energy excitations. These occur when the momentum indices stay a finite
distance from $M/4$ or from $3M/4$ in the limit. The energies created by
$B^\dagger_r$ are 
\bea
\cos\frac{2\pi r}{M}&=&\cos\left(\frac{\pi}{2}+\frac{2\pi(r-M/4)}{M}\right)
=-\sin\frac{2\pi(r-M/4)}{M}\sim -\frac{2\pi(r-M/4)}{M}\nonumber\\
\cos\frac{2\pi s}{M}&=&\cos\left(\frac{3\pi}{2}+\frac{2\pi(s-3M/4)}{M}\right)
=\sin\frac{2\pi(s-M/4)}{M}\sim \frac{2\pi(s-3M/4)}{M}
\eea
Notice that in the limit the excitation energies are plus or minus the
momenta relative to the appropriate Fermi momentum $M/4$ or $3M/4$. Also
keep in mind that $r$ must be chosen half odd or integer according to
whether $\Omega$ of the state is positive or negative. Write the excitation
energies as $2\pi\lambda_r/M$ we see that $\lambda_r$ is integer or half
odd integer when $M$ is even, but is $\pm 1/4+$ integer when $M$ is odd.

The large $M$ behaviors of the ground energies (last column of the table)
are given by:
\bea
-\csc\frac{\pi}{M}&\sim&-\frac{M}{\pi}\left[1+\frac{\pi^2}{6M^2}
  +O(M^{-4})\right]=-\frac{M}{\pi}-\frac{\pi}{6M} +O(M^{-3})\\
-\cot\frac{\pi}{M}&\sim&
 -\frac{M}{\pi}-\frac{\pi}{6M}+\frac{\pi}{2M} +O(M^{-3})\\
-\cot\frac{\pi}{M}\cos\frac{\pi}{2M}&\sim&
-\frac{M}{\pi}-\frac{\pi}{6M}+\frac{\pi}{2M}+\frac{\pi}{8M} +O(M^{-3})
\eea
Inspection shows that the right sides of these equations can be summarized by
\bea
E_G(Q,{\hat P})=-\frac{M}{\pi}-\frac{\pi}{6M}+\frac{\pi}{M}\left[
  \frac{Q^2}{8}+2\frac{{\hat P}^2}{\pi^2}\right]+O(M^{-3})
\eea
which is the known formula for the ground state in the given charge and
momentum sectors. The negative first $1/M$ term is the familiar zero point
energy associated with the closed bosonic string, The ground states in
the higher $Q,{\hat P}$ sectors can be easily constructed by applying
monomials of suitable eigenoperators to the states of the table,
which change to the desired values of $Q,{\hat P}$ creating the lowest
possible amount of energy. For example the operator $B^\dagger_{(M-2)/4}
B^\dagger_{(3M+2)/4}$ adds (-4) to $Q$ and 0 to ${\hat P}$ and $2\pi/M
=4^2\pi/(8M)$ to the energy.
\subsection{Energy levels and degeneracies for $M$ large}
Because the system is one of free fermions it is straightforward to read
off the energy levels and work out their degeneracies. However, it is
useful to organize this information in terms of generating functions,
especially when $M$ is large, when the generating functions are elliptic
functions. There is a famous identity which codes the known possibility
(``bosonization'') of describing free fermion fields
by free compactified bosonic fields. We quote it in the form\footnote{
  In this subsection
  we follow standard conventions and denote the
  elliptic function modulus by $q$. suspending
  the use of $q$ for the number of overturned spins.}
\bea
q^{\alpha^2/4}\prod_{n=1}^\infty(1+e^{2i\theta }q^{2n+\alpha-1})
(1+e^{-2i\theta }q^{2n-\alpha-1})&=&\prod_{n=1}^\infty(1-q^{2n})^{-1}
\sum_{m=-\infty}^\infty q^{(m+\alpha/2)^2}e^{2im\theta}
\eea
Putting $q=e^{-\pi/M}$ and expansing in powers of $q$ generates the energy
spectrum, with the exponents giving the energies and the coefficients
the degeneracy of each level.
Inspecting the large $M$ energies we see we shall need $\alpha=0,1$
for even $M$, and $\alpha=1/2$ for odd $M$. The left side of the identity
describes the spectrum in terms of free fermions and the right
side as bosonic oscillators.

Let's start with the case where $M\in 4\mathbb{N}$. The operator $B^\dagger_r$
is a creation operator for $r\notin G$. It creates low energy for
$r$ near $3M/4$ or $M/4$. In the first case we call the operator $b^\dagger_r$,
which creates $-2$ units of $Q$, $1/2$ unit of ${\hat P}/\pi$, and an odd
  multiple of $\pi/M$ in energy. It yields a factor $e^{-2i\theta
    +i\phi/2} q^{2n-1}$. In the second case we call the operator
  ${\tilde b}^\dagger_r$, which creates $-2$ units of $Q$,
  $-1/2$ unit of ${\hat P}/\pi$, and an odd
  multiple of $\pi/M$ in energy. It yields a factor $e^{-2i\theta
    -i\phi/2} q^{2n-1}$. When $r\in G$, $B_r$ is the creation operator.
  If $r$ is near $3M/4$, it is called $b_r$, and creates
  $2$ units of $Q$, $-1/2$ unit of ${\hat P}/\pi$, and an odd
  multiple of $\pi/M$ in energy. It yields a factor $e^{2i\theta
    -i\phi/2} q^{2n-1}$. Finally the case $ r$ near $M/4$ yields the
  factor $e^{2i\theta +i\phi/2} q^{2n-1}$. Thus the generating function
  in fermion language is
  \bea
\prod_{n=1}^\infty(1+e^{-2i\theta+i\phi/2 }q^{2n-1})
(1+e^{-2i\theta-i\phi/2 }q^{2n-1})(1+e^{2i\theta-i\phi/2 }q^{2n-1})
(1+e^{2i\theta+i\phi/2 }q^{2n-1})
\eea
Only the terms corresponding to an even number of operators applied
to the ground state are relevant, so this formula should be projected onto
that sector. This can be done by symmetrizing it under $\theta\to\theta
+\pi/2$. If we apply the elliptic function identity (with $\alpha=0$)
twice we arrive at an expression
for which the desired projection is transparent. the right side of the
identity is
\bea
\prod_n(1-q^{2n})^{-2}\sum_{k,m} q^{k^2+m^2}e^{2i(k+m)\theta+i(k-m)\phi/2}
  \eea
  In this form the projection is simply to restrict the sum to $k+m$ even.
  To do this call $k+m=K$, so that $k-m=K-2m$. If $K$ is even we can define
  $J=m-K/2$. Then
  \bea
  m^2+k^2=(J+K/2)^2+(J-K/2)^2=2J^2+\frac{K^2}{2}
  \eea
  and the projected expression is
  \bea
  \prod_n(1-q^{2n})^{-2}\sum_{K even,J} q^{K^2/2+2J^2}
  e^{2iK\theta-iJ\phi}
  \eea
  From the way $\theta,\phi$ were introduced, it is clear that
  $J=-{\hat P}/\pi$ and $2K=Q$, so the exponent of $q$ reads $Q^2/8
  +2{\hat P}^2/\pi^2$, the known result. But this sector only includes
  charges $Q\in4\mathbb{Z}$. the odd multiples of 2 are covered by the
  $\Omega=-1$ sector built on the second line, which we turn to next.

  The state described by the second line of the table has $Q=+2$,
  ${\hat P}=0$, and energy $\pi/(2M)$ above the energy of line 1.
  Also the excitation energies are even integers times $\pi/M$.
  It is then straightforward to establish the generating function:
  \bea
&&  e^{2i\theta} q^{1/2}(1+e^{-2i\theta+i\phi/2})(1+e^{-2i\theta-i\phi/2})
\nonumber\\
&&\prod_{n=1}^\infty(1+e^{-2i\theta+i\phi/2 }q^{2n})
(1+e^{-2i\theta-i\phi/2 }q^{2n})(1+e^{2i\theta-i\phi/2 }q^{2n})
(1+e^{2i\theta+i\phi/2 }q^{2n})\nonumber\\
&&= q^{1/2}(e^{i\theta-i\phi/4}+e^{-i\theta+i\phi/4})
(e^{i\theta+i\phi/4}+e^{-i\theta-i\phi/4}))
\nonumber\\
&&\prod_{n=1}^\infty(1+e^{-2i\theta+i\phi/2 }q^{2n})
(1+e^{-2i\theta-i\phi/2 }q^{2n})(1+e^{2i\theta-i\phi/2 }q^{2n})
(1+e^{2i\theta+i\phi/2 }q^{2n})
\eea
Of course only the terms corresponding to an even number of excitation
above the second line ground state are relevant. These are singled out by antisymmetrizing the whole expression under $\theta\to\theta+\pi/2$. Again
we can apply the elliptic function (with $\alpha=1$) twice to rewrite
the generating function
\bea
\prod_n(1-q^{2n})^{-2}\sum_{k,m} q^{(k+1/2)^2+(m+1/2)^2}
e^{2i(k+m+1)\theta+i(k-m)\phi/2}
\eea
The projection is achieved by restricting the $k,m$ sum to $k+m$ even.
in terms of the $J,K$ summation indices introduced previously the generating
function reads
\bea
\prod_n(1-q^{2n})^{-2}\sum_{K even,J} q^{(K+1)^2/2+2J^2}
e^{2i(K+1)\theta-iJ\phi}
=\prod_n(1-q^{2n})^{-2}\sum_{K odd,J} q^{K^2/2+2J^2}
e^{2iK\theta-iJ\phi}
\eea
Putting the two sectors together simply relaxes the even or odd restriction:
\bea
\prod_n(1-q^{2n})^{-2}\sum_{K,J} q^{K^2/2+2J^2}
e^{2iK\theta-iJ\phi},\qquad Q\in 2\mathbb{Z}.
\eea
The analysis for $M\in2+4\mathbb{n}$ is the same in spite of the flip
in $\Omega$ assignments because the momenta relative to the Fermi
momenta is identical in both cases.

The analysis for odd $M$ is a bit more involved. Starting with the
ground state described by line 7 of the table, we read off the
generating function:
  \bea
&&  e^{i\theta-i\phi/2} q^{5/8}
\nonumber\\
&&\prod_{n=0}^\infty(1+e^{-2i\theta+i\phi/2 }q^{2n-1/2})
(1+e^{-2i\theta-i\phi/2 }q^{2n+3/2})(1+e^{2i\theta-i\phi/2 }q^{2n+5/2})
(1+e^{2i\theta+i\phi/2 }q^{2n+1/2})\nonumber\\  
&&=  e^{i\theta-i\phi/2} q^{5/8}\prod_{n=1}^\infty(1+e^{2i\theta+i\phi/2 }q^{2n-3/2})(1+e^{-2i\theta-i\phi/2 }q^{2n=1/2})
\nonumber\\
&&\prod_{n=1}^\infty(1+e^{-2i\theta+i\phi/2 }q^{2n-5/2})
(1+e^{2i\theta-i\phi/2 }q^{2n+1/2})
\eea
We use the elliptic function identities for $\alpha=-1/2$ and
$\alpha=3/2$ to arrive at the bosonic form
\bea
 &&e^{i\theta-i\phi/2}\prod_n(1-q^{2n})^{-2}\sum_{k,m} q^{(k-1/4)^2+(m+3/4)^2}
 e^{2i(k+m+1)\theta+i(k-m)\phi/2}\nonumber\\
 &&=\prod_n(1-q^{2n})^{-2}\sum_{k,m} q^{(k-1/4)^2+(m+3/4)^2}
 e^{i(2(k+m)+1)\theta+i(k-m-1)\phi/2}
\eea
Again only the terms with $k+m$ even are relevant. With this restriction
the change of indices $k=K/2-J$, $m=K/2+J$ is valid, so
\bea
(k-1/4)^2+(m+3/4)^2&=&(K/2-J-1/4)^2+(K/2+J+3/4)^2\nonumber\\
&=&2(K/2+1/4)^2+2(J+1/2)^2
\eea
yielding the result
\bea
\prod_n(1-q^{2n})^{-2}\sum_{K even,J} q^{(2K+1)^2/8+2(J+1/2)^2}
 e^{i(2K+1)\theta-i(J+1/2)\phi}
 \eea
 Considering the states built on line 9 of the table, shows that
 the generating function for such excited states is the complex conjugate
 of the one just computed for line 7:
 \bea
&&\hskip-.25in\prod_n(1-q^{2n})^{-2}\sum_{K even,J} q^{(2K+1)^2/8+2(J+1/2)^2}
e^{-i(2K+1)\theta+i(J+1/2)\phi}\nonumber\\
&&=\prod_n(1-q^{2n})^{-2}\sum_{K even,J} q^{(2K-1)^2/8+2(J-1/2)^2}
e^{i(2K-1)\theta-i(J-1/2)\phi}\nonumber\\
&&=\prod_n(1-q^{2n})^{-2}\sum_{K even,J} q^{(2(K-1)+1)^2/8+2(J+1/2)^2}
e^{i(2(K-1)+1)\theta-i(J+1/2)\phi}\nonumber\\
&&=\prod_n(1-q^{2n})^{-2}\sum_{K odd,J} q^{(2K+1)^2/8+2(J+1/2)^2}
e^{i(2K+1)\theta-i(J+1/2)\phi}
\eea
In successive lines we have reversed the signs of $K,J$, Shifted $J$ by one
unit, and shifted $K$ by one unit making it odd rather than even,
Thus putting the two sectors together simply relaxes the restrictions on the
range of $K$,
\bea
\prod_n(1-q^{2n})^{-2}\sum_{K,J} q^{(2K+1)^2/8+2(J+1/2)^2}
e^{i(2K+1)\theta-i(J+1/2)\phi},\qquad Q\in 1+2\mathbb{Z}
\eea
In spite of the need to allow for 8 base ground states, only four lead to
distinct outcomes for energies and degeneracies.
They can be characterized by the nature of $Q$:
\bea
Q\in 4\mathbb{Z},\qquad Q\in 2+ 4\mathbb{Z},\qquad Q\in 1+ 4\mathbb{Z},
\qquad  Q\in -1 +  4\mathbb{Z}
\eea

\section{Conformal Invariance and Bosonization at Large $M$}
\subsection{Virasoro generators}
We can summarize the  spectrum using energy eigenoperators by writing
\bea
H&=&\sum_r \cos\frac{2\pi r}{M}B^\dagger_r B_r\\
&=&E^\Omega_G+\sum_{r\notin G} \cos\frac{2\pi r}{M}B^\dagger_r B_r
-\sum_{r\in G} \cos\frac{2\pi r}{M}B_r B^\dagger_r
\eea
And we remember that $r\in 1/2+\mathbb{Z}$ when $\Omega=+1$ , and
$r\in \mathbb{Z}$ when $\Omega=-1$.
When $M$ is large $B^\dagger_r$ creates energies of order $1/M$ for
$r$ near $3M/4$ or near $M/4$. In the first case we call the operator
$b^\dagger_{r-3M/4}$ and in the second case ${\tilde b}^\dagger_{r-M/4}$.
Then we can write the  effective low energy Hamiltonian
\bea
H_{\rm eff}&=&E^\Omega_G+\frac{2\pi}{M}\left[\sum_{s\notin G} sb^\dagger_s b_s
  +\sum_{t\notin G} t{\tilde b}^\dagger_t {\tilde b}_t+\sum_{u\in G}
(-u)b_u b^\dagger_u+\sum_{v\in G}
(-v){\tilde b}_v {\tilde b}^\dagger_v\right].
\eea
We have written each summation index as a different letter, because their
ranges can be different depending on the nature of $Q$:
\bea
Q&\in& 4\mathbb{Z}:\quad s,t,u,v\in \frac{1}{2}+\mathbb{Z},\qquad s,t>0,
\quad u,v<0\\
Q&\in& 2+4\mathbb{Z}:\quad s,t,u,v\in \mathbb{Z},\qquad s,t\geq 0,
\quad u,v<0\\
Q&\in&1+ 4\mathbb{Z}:\quad s,t,u,v\in -\frac{1}{4}+\mathbb{Z},\qquad s\geq 
-\frac{1}{4},\quad t\geq\frac{3}{4},
\quad u\leq -\frac{5}{4},\quad v\leq-\frac{1}{4}\\
Q&\in& -1+4\mathbb{Z}:\quad s,t,u,v\in \frac{1}{4}+ \mathbb{Z},\qquad  s\geq
\frac{5}{4},\quad t\geq\frac{1}{4},
\quad u\leq \frac{1}{4},\quad v\leq-\frac{3}{4}
\eea
We list explicitly the Hamiltonian for each case, dropping the
term in $E_G$ linear in $M$:
\bea
H_{\rm eff}&=&-\frac{\pi}{6M}+\frac{2\pi}{M}\sum_{n=0}^\infty
\left[\left(n+\frac{1}{2}\right)b^\dagger_{n+1/2}
  b_{n+1/2}
  + \left(n+\frac{1}{2}\right){\tilde b}^\dagger_{n+1/2}{\tilde b}_{n+1/2}
\right]\nonumber\\
&&\hskip-.5in  +\frac{2\pi}{M}\sum_{n=0}^\infty
\left[\left(n+\frac{1}{2}\right)b_{-n-1/2} b^\dagger_{-n-1/2}
  +\left(n+\frac{1}{2}\right){\tilde b}_{-n-1/2} {\tilde b}^\dagger_{-n-1/2}
\right],\quad Q\in 4\mathbb{Z}\\
H_{\rm eff}&=&-\frac{\pi}{6M}+\frac{\pi}{2M}
+\frac{2\pi}{M}\sum_{n=1}^\infty
n\left[b^\dagger_{n}
  b_{n}
  + {\tilde b}^\dagger_{n}{\tilde b}_{n}+
  b_{-n} b^\dagger_{-n}
  +{\tilde b}_{-n} {\tilde b}^\dagger_{-n}
\right],\quad Q\in 2 + 4\mathbb{Z}\\
H_{\rm eff}&=&-\frac{\pi}{6M}+\frac{5\pi}{8M}+\frac{2\pi}{M}\sum_{n=0}^\infty
\left[\left(n-\frac{1}{4}\right)b^\dagger_{n-1/4}
  b_{n-1/4}
  + \left(n+\frac{3}{4}\right){\tilde b}^\dagger_{n+3/4}{\tilde b}_{n+3/4}
\right]\nonumber\\
&&\hskip-.5in  +\frac{2\pi}{M}\sum_{n=0}^\infty
\left[\left(n+\frac{5}{4}\right)b_{-n-5/4} b^\dagger_{-n-5/4}
  +\left(n+\frac{1}{4}\right){\tilde b}_{-n-1/4} {\tilde b}^\dagger_{-n-1/4}
\right],\quad Q\in 1+4\mathbb{Z}\\
H_{\rm eff}&=&-\frac{\pi}{6M}+\frac{5\pi}{8M}+\frac{2\pi}{M}\sum_{n=0}^\infty
\left[\left(n+\frac{5}{4}\right)b^\dagger_{n+5/4}
  b_{n+5/4}
  + \left(n+\frac{1}{4}\right){\tilde b}^\dagger_{n+1/4}{\tilde b}_{n+1/4}
\right]\nonumber\\
&&\hskip-.5in  +\frac{2\pi}{M}\sum_{n=0}^\infty
\left[\left(n-\frac{1}{4}\right)b_{-n+1/4} b^\dagger_{-n+1/4}
  +\left(n+\frac{3}{4}\right){\tilde b}_{-n-3/4} {\tilde b}^\dagger_{-n-3/4}
\right],\quad Q\in -1+4\mathbb{Z}
\eea
The operator products in the above expressions are all normal ordered
relative to the ground state in each sector, that is operators which annihilate
the ground state are on the right. If we understand the double colon
notation to mean normal ordering in this sense, we can give these expressions
more compact forms:
\bea
H^0_{\rm eff}&=&-\frac{\pi}{6M}+\frac{2\pi}{M}\sum_{n=-\infty}^\infty
\left(n+\frac{1}{2}\right)\left[:b^\dagger_{n+1/2}
  b_{n+1/2}:
  +:{\tilde b}^\dagger_{n+1/2}{\tilde b}_{n+1/2}:
\right]\nonumber\\
H^2_{\rm eff}&=&-\frac{\pi}{6M}+\frac{\pi}{2M}
+\frac{2\pi}{M}\sum_{n=-\infty}^\infty
n\left[:b^\dagger_{n}b_{n}:
  + :{\tilde b}^\dagger_{n}{\tilde b}_{n}:
\right]\nonumber\\
H^1_{\rm eff}&=&-\frac{\pi}{6M}+\frac{5\pi}{8M}
+\frac{2\pi}{M}\sum_{n=-\infty}^\infty
\left(n-\frac{1}{4}\right)\left[:b^\dagger_{n-1/4}
  b_{n-1/4}:
  + :{\tilde b}^\dagger_{n-1/4}{\tilde b}_{n-1/4}:
\right]\nonumber\\
H^{-1}_{\rm eff}&=&-\frac{\pi}{6M}+\frac{5\pi}{8M}+\frac{2\pi}{M}
\sum_{n=-\infty}^\infty
\left(n+\frac{1}{4}\right)
\left[:b^\dagger_{n+1/4}
  b_{n+1/4}:
  + :{\tilde b}^\dagger_{n+1/4}{\tilde b}_{n+1/4}:
\right]\nonumber
\eea
In each sector, the Hamiltonian is a sum of two commuting operators,
one depending only on $b$ and the other on ${\tilde b}$. Each can be
identified with the zero member of commuting Virasoro generators,
$L_n$, ${\tilde L}_n$. The Virasoro algebras must take the form
\bea
{}[L_n,L_m]=(n-m)L_{n+m}+ C_n\delta_{n,-m}
\eea
and similarly for ${\tilde L}_n$. When $n\neq0$, there is no ordering
ambiguity in the definition of $L_n$ so allowing for fractional modes
we may write
\bea
L_n\equiv \sum_{k=-\infty}^\infty
(k+n/2+\alpha)b^\dagger_{k+\alpha}b_{k+n+\alpha},\qquad n\neq0
\eea
This expression is ambiguous up to an additive constant for $n=0$.
But $L_0$ can be obtained using the algebra of
the unambiguous $L_n$: namely $[L_n,L_{-n}]=2nL_0+C_n$, once conventions
are set in determining $C_n$. Those conventions will be set by identifying
the $L_0, {\tilde L}_0$ with pieces of the Hamiltonians in the various sectors.
Inspection shows that the appropriate values of $\alpha$ are $1/2,0,1/4,-1/4$
for the respective sectors.

It is easy to confirm the Virasoro
algebra for $n\neq -m$, but the determination of the $C_n$ is more involved.
\bea
L_0^0&=&\sum_{n=0}^\infty\left(n+\frac{1}{2}\right)
\left[b^\dagger_{n+1/2}
  b_{n+1/2}+b_{-n-1/2} b^\dagger_{-n-1/2}\right]\nonumber\\
{\tilde L}_0^0&=&\sum_{n=0}^\infty\left(n+\frac{1}{2}\right)
\left[{\tilde b}^\dagger_{n+1/2}
  {\tilde b}_{n+1/2}+{\tilde b}_{-n-1/2} {\tilde b}^\dagger_{-n-1/2}\right]\\
L_0^2&=&\sum_{n=0}^\infty\left(n\right)
\left[b^\dagger_{n}
  b_{n}+b_{-n} b^\dagger_{-n}\right],\qquad
{\tilde L}_0^0=\sum_{n=0}^\infty\left(n\right)
\left[{\tilde b}^\dagger_{n}
  {\tilde b}_{n}+{\tilde b}_{-n} {\tilde b}^\dagger_{-n}\right]
\eea
\bea
L_0^{1}&=&\sum_{n=0}^\infty
\left[\left(n-\frac{1}{4}\right)b^\dagger_{n-1/4}
  b_{n-1/4}+\left(n+\frac{5}{4}\right)b_{-n-5/4} b^\dagger_{-n-5/4}\right]
  \nonumber\\
  {\tilde L}_0^1&=&\sum_{n=0}^\infty
  \left[ \left(n+\frac{3}{4}\right){\tilde b}^\dagger_{n+3/4}{\tilde b}_{n+3/4}
  +\left(n+\frac{1}{4}\right){\tilde b}_{-n-1/4} {\tilde b}^\dagger_{-n-1/4}
\right]\\
L_0^{-1}&=&\sum_{n=0}^\infty
\left[\left(n+\frac{5}{4}\right)b^\dagger_{n+5/4}
  b_{n+5/4}+\left(n-\frac{1}{4}\right)b_{-n+1/4} b^\dagger_{-n+1/4}\right]
  \nonumber\\
  {\tilde L}_0^{-1}&=&\sum_{n=0}^\infty
  \left[ \left(n+\frac{1}{4}\right){\tilde b}^\dagger_{n+1/4}
    {\tilde b}_{n+1/4}
  +\left(n+\frac{3}{4}\right){\tilde b}_{-n-3/4} {\tilde b}^\dagger_{-n-3/4}
\right]
\eea
For each of the four cases, we have 2 commuting Virasoro algebras, with
potentially different central terms for each. With the above identifications
for $L_0$ in each case, careful calculation shows
\bea
C_n^0&=&{\tilde C}_n^0=\frac{1}{12}(n^3-n)\nonumber\\
C_n^2&=&{\tilde C}_n^2=\frac{1}{12}(n^3+2n)\nonumber\\
C_n^1&=&\frac{1}{12}(n^3+23n/4),\qquad
{\tilde C}_n^1=\frac{1}{12}(n^3-n/4)\nonumber\\
C_n^{-1}&=&\frac{1}{12}(n^3+23n/4),\qquad{\tilde C}_n^{-1}=
\frac{1}{12}(n^3-n/4)
\eea
The coefficient of $n^3$ is universal corresponding to a single bosonic
worldsheet coordinate. This term is controlled by short distances and is
expected to be independent of the structure of the ground state. In contrast,
the linear term varies from one case to another. It is perhaps more
natural to include the linear term in the definition of $L_0$, keeping
only the cubic term in the definition of $C_n$. If this is done one finds
that the constant terms in the Hamiltonian are precisely accounted for
with the simple identification $H=2\pi(L_0+{\tilde L}_0)/M$.

The asymmetry between $C$ and ${\tilde C}$ in the cases that $Q$ is odd,
though a little startling, is understandable given that the occupied states
in the chosen vacuum are asymmetric. As seen in the table the ground
energy level  with 
$Q=1$ is doubly degenerate, and choosing the other ground state
reverses the asymmetry. 
\subsection{Bosonization of the spin system at $\Delta=0$}
The equivalence of free fermion systems to bosonic systems in one space
dimension is very well studied. In this subsection we quote the equivalence
in the operator language for the Heisenberg spin system studied
in this article. There are four sectors, each of which is described by
two anticommuting sets of fermion operators $b, {\tilde b}$, so all together
eight slightly different bosonization formulas. Each set is characterized by
a fractional mode number, which we have called $\alpha$. We first quote
the formulas for general $\alpha$, and use $b$ for the Fermi description,
and $a$ for the bosonic description.

We write the boson as a bilinear of fermions
\bea
a_n&=&\sum_{k=-\infty}^\infty : b^\dagger_{k+\alpha}b_{k+n+\alpha}:
\eea
where we fix the normal ordering with respect to a vacuum satisfying
$b_{k+\alpha}\ket{0}=0$ for $k\geq0$ and $b^\dagger_{k+\alpha}
\ket{0}=0$ for
$k<0$.  it follows from  this formula
and the anticommutation
relations $\{b_{k+\alpha}, b^\dagger_{l+\alpha}\}=\delta_{kl}$
 that
\bea
{}[a_n,a_m]=n\delta_{n,-m},\qquad a_n^\dagger=a_{-n},\qquad a_n\ket{0}=0,
\quad n\geq 0.
\eea
When calculating the commutator care must be taken to respect normal
ordering: each term in the commutator should be separately normal ordered
via Wick;s theorem, after which the operator parts cancel leaving behind the
right side of the equation.

Bosonization in one space dimension, remarkably, leads to an equality
between two relatively simple Hamiltonians expressed in terms of bosonic
or fermionic degrees of freedom. For this system at the free fermion point
this relation holds for the entire Virasoro algebra. The fermion form of
the Virasoro algebra was given in the previous subsection. The bosonic
form is given by\footnote{More generally one could also add a term
  $in\gamma a_n$ to $L_n$ which modifies the cubic term in $C_n$,
but since the
two systems already agree on the cubic term, $\gamma=0$ for us.}
\bea
L_n&=&\beta a_n+\frac{1}{2}\sum_{m=-\infty}^\infty : a_{-m}a_{m+n}:
\eea
Here normal ordering simply means positive moded $a$'s always stand
to the right of negative moded $a$'s. Since we will be inserting the
bosonization formula, which uses its own normal ordering prescription,
we write out $L_n$ explicitly:
\bea
L_n&=&\beta a_n+\frac{1}{2}\sum_{m=0}^\infty a_{-m}a_{m+n}
+\frac{1}{2}\sum_{m=-\infty}^{-1} a_{m+n}a_{-m}\nonumber\\
&=&\beta a_n+\frac{1}{2}\sum_{m=0}^\infty a_{-m}a_{m+n}
+\frac{1}{2}\sum_{m=1}^{\infty} a_{-m+n}a_{m}
\eea
so that normal ordering is no longer needed.

Now one can plug the bosonization formula into $L_n(a)$ and, with
due attention to operator ordering,  obtain
\bea
L_n(a)&=& \sum_{k=-\infty}^\infty
(k+n/2+\beta+1/2):b^\dagger_{k+\alpha}b_{k+n+\alpha}:
\eea
Comparison gives $\beta=\alpha-1/2$.

In bosonic language the $\beta=\alpha-1/2$ term in $L_n$ is equivalent to
the replacement $a_0\to a_0+\beta$ for $n\neq0$. For $n=0$ the
$a_0$ dependence is
\bea
\frac{a_0^2}{2}+\beta a_0=\frac{(a_0+\beta)^2}{2}-\frac{\beta^2}{2}
\eea
so in addition to the substitution $a_0\to a_0+\beta$ a constant term
added to $L_0$ appears.
With this definition of $L_0$ the c-number term in the Virasoro algebra
assumes the form $C_n=\beta^2+(n^3-n)/12$. Clearly if one removes that extra term from $L_0$ and puts it in the the c-number term, he latter becomes
$C_n\to (n^3-n)/12$, the value it has at $\beta=0$. In short, with the new
definitions
\bea
a_0&=&\sum_{k=-\infty}^\infty : b^\dagger_{k+\alpha}b_{k+\alpha}+
\alpha-\frac{1}{2}:\\
L_n&=&\sum_{k=-\infty}^\infty
(k+n/2+\alpha):b^\dagger_{k+\alpha}b_{k+n+\alpha}:
+\frac{1}{2}\left(\alpha-\frac{1}{2}\right)^2\delta_{n0},
\eea
the $L_n$, whose bosonic form is $L_n=\sum_m:a_{-m}a_{m+n}:/2$,
satisfy the standard Virasoro algebra for transverse dimension $d=1$
\bea
{}[L_n,L_m]=(n-m)L_{n+m}+ \frac{1}{12}(n^3-n)\delta_{n,-m}
\eea
\subsection{Interpretation of zero mode operators} 
The zero modes $a_0$, ${\tilde a}_0$ can be linked to the two conserved
operators $Q$ and ${\hat P}/\pi$, whose values for the ground states
are listed in the table:
\bea
Q^0&=&-2\sum_{n=0}^\infty\left[:b^\dagger_{n+1/2}b_{n+1/2}:
  + :{\tilde b}^\dagger_{n+1/2}{\tilde b}_{n+1/2}: \right]=-2(a_0+{\tilde a}_0)
\nonumber\\
\frac{{\hat P}^0}{\pi}&=&\frac{!}{2}\sum_{n=0}^\infty\left[:b^\dagger_{n+1/2}
  b_{n+1/2}:
  - `:{\tilde b}^\dagger_{n+1/2}{\tilde b}_{n+1/2}: \right]
=\frac{1}{2}(a_0-{\tilde a}_0)
\\
Q^2&=&2-2\sum_{n=0}^\infty\left[:b^\dagger_{n}b_{n}:
+ :{\tilde b}^\dagger_{n}{\tilde b}_{n}:\right]=-2(a_0+{\tilde a}_0)
\nonumber\\
\frac{{\hat P}^2}{\pi}&=&\frac{!}{2}\sum_{n=0}^\infty\left[:b^\dagger_{n}
  b_{n}:  - :{\tilde b}^\dagger_{n}{\tilde b}_{n}: \right]
=\frac{1}{2}(a_0-{\tilde a}_0)
\eea
\bea
Q^1&=&1-2\sum_{n=0}^\infty\left[:b^\dagger_{n-1/4}b_{n-1/4}:
+ :{\tilde b}^\dagger_{n+3/4}{\tilde b}_{n+3/4}: \right]=-2(a_0+{\tilde a}_0)
\nonumber\\
\frac{{\hat P}^1}{\pi}&=&-\frac{!}{2}+\frac{!}{2}\sum_{n=0}^\infty
\left[:b^\dagger_{n+3/4}
  b_{n+3/4}:
- :{\tilde b}^\dagger_{n+3/4}{\tilde b}_{n+3/4}: \right]
=\frac{1}{2}(a_0-{\tilde a}_0)
\\
Q^{-1}&=&-1-2\sum_{n=0}^\infty\left[:b^\dagger_{n+5/4}b_{n+5/4}:
+ :{\tilde b}^\dagger_{n+1/4}{\tilde b}_{n+1/4}: \right]=-2(a_0+{\tilde a}_0)
\nonumber\\
\frac{{\hat P}^{-1}}{\pi}&=&\frac{!}{2}+\frac{!}{2}\sum_{n=0}^\infty
\left[:b^\dagger_{n+5/4}
  b_{n+5/4}:
- :{\tilde b}^\dagger_{n+1/4}{\tilde b}_{n+1/4}: \right]
=\frac{1}{2}(a_0-{\tilde a}_0)
\eea
in all cases the new definitions of $a_0$ and ${\tilde a}_0$ are used on the
extreme right side. In a similar vein,using the new definition of $L_0$,
${\tilde L}_0$, the Hamiltonian is given by
\bea
H&=&-\frac{\pi}{6M}+\frac{2\pi}{M}(L_0+{\tilde L}_0)
\eea
Furthermore if  $L_0$, ${\tilde L}_0$ are defined so that the c-number term
in the
Virasoro algebra is purely cubic in $n$, the $-\pi/(6M)$ term on the
right of the above equation should be deleted.
\section{Spin Chain Contribution to 3 String Vertex.}
\subsection{Preliminary Discussion}
Perturbative string interactions are efficiently obtained using worldsheet
  path integrals. In lightcone parameters, using
  Mandelstam's interacting string formalism, 
  any worldsheet can be tessellated by a series of rectangular pieces each
  describing the propagation of a free string. All interactions can be
  built up from the basic three string vertex at which a long string
  whose propagator is a rectangle of width $P_1^++P_2^+$ makes an
  instantaneous transition
  to or from two shorter strings whose propagators have width $P_1^+$
  and $P_2^+$. Factorizing each propagator shows that the vertex is
  nothing more than the overlap of the long string wave function with the
  product of the two short string wave functions, the amplitude for
  the transition in one direction being the complex conjugate of
  the amplitude for the time reversed transition.

  A given string theory is defined by a number of worldsheet fields,
  typically including bosonic coordinates 
compactified or not, Grassmann fields
  representing spin degrees of freedom, or more generally any worldsheet system
  that supports the Virasoro algebra. In our string bit models the worldsheet
  is not fundamental, but rather is a composite structure describing the
  evolution of chains of string bits with time. Any fields living on the
  worldsheet must arise from internal degrees of freedom
  carried by the string bit,
  and indeed need arise from only a finite number of bit degrees of freedom.
  In this paper, we are exploring the possibility that each spin of
  the Heisenberg system resides on a single string bit. Then the Heisenberg
  chain arises when a large number of bits link together. To the extent that
  the effective worldsheet fields arising  in this way are decoupled
  from each other the basic overlap is just a product of overlaps for
  each subsystem. Then we can focus on the overlap of the Heisenberg
  wave functions alone, as we have done in previous sections.

  Away from the free fermion point ($\Delta\neq0$) the
  Hamiltonian has terms up to quartic in fermion fields, making the
  overlap calculation very difficult. However, by bosonizing the
  system, we obtain a Hamiltonian quadratic in boson fields, with
  zero modes quantized according to the eigenvalues of $Q$ and ${\hat P}/\pi$.
  We interpret $Q$ as a Kaluza-Klein momentum of a compactified coordinate
  and ${\hat P}/\pi$ as the associated winding number. Our strategy for
  obtaining the overlap even with $\Delta\neq0$ is to use Mandelstam's
  interacting string formalism for bosonic string, generalized to
  allow for compactified coordinates. The only loose end in this approach,
  is figuring out how to deal with the two zero modes. If all three
  states participating in the overlap have ${\hat P}=0$, it is
  probably safe to use the Mandelstam vertex for continuous (uncompactified)
  momenta by simply setting those momenta equal to the desired KK momentum
  proportional to $Q$. This is because $Q=Q_1+Q_2$ is an operator equality,
  showing that the vertex conserves $Q$.
  The same is not true of the momentum operator, so the possibility that
  the vertex violates ${\hat P}$ conservation can not be ruled out.
  Indeed, This is known to occur for the Green-Schwarz worldsheet
  spinors. We shall find in the next subsection that for the Heisenberg
  spin system ${\hat P}$ is indeed conserved in the string limit.
\subsection{Three string overlap}
    In string theory the interactions of strings arise from instantaneous
    changes in the Hamiltonian, corresponding to the transition of a single
    closed string to two smaller closed strings. In the string bit approach
    this process does not change the number of string bits. The overlap
    involves all available degrees of freedom, but here we only deal with
    those described by Heisenberg spins. The Hamiltonian for any single
    spin chain is the one just described. Let's say that the initial chain
    has $M$ spins, $S_1,\ldots, S_M$ and the smaller chains have $K$ and
    $L=M-K$ spins respectively. We identify $S_1,\ldots, S_K$ as the spin
    variables of the first small chain and $S_{K+1},\ldots,S_M$ as those of
    the second small chain,

    We construct the raising and lowering operators for the three participating
    chains
    \bea
B_r&=&\frac{1}{\sqrt{M}}\sum_{k=1}^M e^{-2\pi ikr/M}S_k,\qquad 0\leq r<M\\
b^1_r&=&\frac{1}{\sqrt{K}}\sum_{k=1}^K e^{-2\pi ikr/K}S_k,\qquad 0\leq r<K\\
b^2_r&=&\frac{1}{\sqrt{L}}\sum_{k=K+1}^M e^{-2\pi i(k-K)r/L}S_k,
\qquad 0\leq r<L
    \eea
    To apply  the overlap construction developed in Section 2,
    we need the matrix $D=(D^1|D^2)$
    \bea
    D^1_{sr}&=&\{b^1_r,B^\dagger_s\}=\frac{1}{\sqrt{KM}}
    \sum_{k=1}^K e^{2\pi ik(s/M-r/K)}\nonumber\\
    &=&\begin{cases}\sqrt{\frac{\displaystyle{K}}{\displaystyle{M}}}, &
      (s/M-r/K)\in \mathbb{Z}\\
-\frac{\displaystyle1}{\displaystyle\sqrt{KM}}
\frac{\displaystyle{1-e^{2\pi i(sK/M-r)}}}{\displaystyle{
    1-e^{-2\pi i(s/M-r/K)}}},    &(s/M-r/K)\notin
\mathbb{Z}\end{cases}\\
D^2_{sr}&=&\{b^2_r,B^\dagger_s\}=\frac{1}{\sqrt{LM}}
    \sum_{k=K+1}^M e^{2\pi i(ks/M-(k-K)r/L)}=\frac{1}{\sqrt{LM}}
    e^{2\pi iKs/M}\sum_{k=1}^L e^{2\pi i(ks/M-kr/L)}\nonumber\\
    &=&e^{2\pi iKs/M}\begin{cases}\sqrt{\frac{\displaystyle{L}}{\displaystyle{
            M}}}, &
      (s/M-r/L)\in \mathbb{Z}\\
-\frac{1}{\sqrt{\displaystyle{LM}}}
\frac{\displaystyle{1-e^{2\pi i(sL/M-r)}}}{\displaystyle{
    1-e^{-2\pi i(s/M-r/L)}}},
&(s/M-r/L)\notin \mathbb{Z}\end{cases}    \eea
The special cases can easily be seen to follow from a limit
of the generic cases, assuming $K,L,M$ are continuous variables.
It is only necessary to treat them separately in evaluations
(e.g. on a computer) which explicitly require $K,L,M$ to be integers.

We also need a common energy eigenstate, which we called $\ket{0}$ in 
Section 2. A convenient choice is the state in which all spins
are up, $\sigma_k^z\ket{0}=\ket{0}$. This state is an eigenstate of $\Omega$
with value $+1$ and satisfies $S_k\ket{0}=0$, $B_k\ket{0}=0$,
$b^1_k\ket{0}=0$, and $b^2_k\ket{0}=0$, for all $k$. It is also an
eigenstate of the Hamiltonian, even for $\Delta\neq0$
\bea
H_\Delta\ket{0}=\Delta\sum_k\sigma_k^z\sigma_{k+1}^z\ket{0}=M\Delta\ket{0}
\eea
We shall refer to $\ket{0}$ as the ``empty'' state. It is {\it not} the
ground state of either system because there are negative energy levels
in both.

Energy eigenstates can be constructed by
applying monomials of the $B^\dagger$'s
or monomials of $b^{1\dagger},b^{2\dagger}$'s to the empty state,
In the first case one builds energy eigenstates of $H_1$
and in the second case those of
$H_2$, Even monomials give eigenstates with $\Omega=+1$ and odd monomials give
eigenstates with $\Omega=-1$. In the case of $H_1$ which describes a single
spin chain, the operators that go into the even monomials
carry half odd integer indices and those going into the odd monomials integer
indices. In the case of $H_2$, which describes two spin chains with
independent operators for each,
the operators that go into the even monomials
either have half odd integer indices for both chains or integer indices for
both chains, and those going into the odd monomials have half odd
integer indices for either of the two chains but integer
indices for the other one.

In order to make efficient use of the coherent state formalism, we consider
two distinct Hamiltonians for each chain: $H^+$ for which $\Omega$
has been set equal to $+1$ and $H^-$ for which $\Omega$ has
been set equal to $-1$. Eigenstates of $H^\pm$ with $\Omega=\mp1$ are not
eigenstates of $H$ and must be discarded.
But eigenstates of $H^\pm$ with $\Omega=\pm1$ {\it are} eigenstates of $H$,
and together span the eigenspace.
For the three chain vertex, one does four separate calculations for the
Hamiltonian choices $+\to++$, $+\to--$, $-\to +-$, and $-\to -+$. At the end
of the calculations, one discards the monomial choices that violate the
required correlation between $\Omega$ value and index.

In Section 2 we gave the generic construction of the  vertex for
systems of free fermions summarized in (\ref{ffoverlap}). We need to form four
(not necessarily square) matrices from $D_{sr}$, where $s$ labels the
modes of the long chain, and $r$ labels the modes of the two short strings:
\bea
D^{Gg}_{sr},\qquad D^{{\tilde G}g}_{sr},\qquad D^{G{\tilde g}}_{sr},
\qquad D^{{\tilde G}{\tilde g}}_{sr},
\eea
Here the superscript $G$ means $s$ labels the levels filled in $G$,
${\tilde G}$ means $s$ labels the levels that are empty in $G$, with a
similar meaning for the superscripts $g,{\tilde g}$ with respect to the $r$
index. Since the overlap construction requires the inverse of the first
of these 4 matrices, it is important that $D^{Gg}$ be a square matrix\footnote{
  In a situation where it is not, modification would be necessary.}.
Inspection of our table (\ref{groundmomenta}) in all of the numerous allowed
cases confirms this to be true. To illustrate consider the cases in which
the bit numbers of all three strings are multiples of 4, so that we can limit
attention to the first three lines of the table. The symbol $q$ is the number
of levels filled in the corresponding ground state. Thus there are
$M/2$ (or $K/2$ or $L/2$) labels for the state described by the first line,
$(M\mp2)/2$ (or $(K\mp2/2$ or $(L\mp2)/2$) for the state described by the
second and third line respectively. One immediately sees that a square
matrix $q=q_1+q_2$ is equivalent to charge conservation $Q=Q_1+Q_2$. Here
we are presuming that $\VEV{G|g}\neq0$.

The 3 chain overlap becomes the three string vertex in the limit that
each chain has an infinite number of bits: $M,K,L=M-K\to \infty$ at
fixed ratio $0<x=K/M<1$.
In this limit the relevant modes on each string are those
with either $r-M_k/4$ of $r-3M_k/4$ fixed. To implement the simplifications
of this stringy limit on the overlap matrices $D^1,D^2$ we have to consider
four cases for each. Define
\bea
s^\prime&\equiv& s-\frac{3M}{4},\qquad {\tilde s}^\prime\equiv s-\frac{M}{4}
\nonumber\\
r^\prime&\equiv& r-\frac{3K}{4},\qquad {\tilde r}^\prime\equiv r-\frac{K}{4}
\nonumber\\
t^\prime&\equiv& t-\frac{3L}{4},\qquad {\tilde t}^\prime\equiv t-\frac{L}{4}
\eea
which are held fixed as $M\to\infty$ Then in the string limit
\bea
D^1_{sr}&\approx&\begin{cases}-{\displaystyle{\sqrt{x}}}
  \frac{\displaystyle{
      1-\exp\{2\pi i({\tilde s}^\prime x-{\tilde r}^\prime )}\}}{
    \displaystyle{2\pi i(
      {\tilde s}^\prime x-{\tilde r}^\prime )}},
  & {\tilde s}^\prime,\; {\tilde r}^\prime\; {\rm fixed}\\
  -{\displaystyle{\sqrt{x}}}
  \frac{\displaystyle{1-\exp\{2\pi i(s^\prime x-r^\prime)}\}}{
    \displaystyle{2\pi i(
      s^\prime x-r^\prime)}},
  & s^\prime,\; r^\prime\; {\rm fixed}\\
  O\left(\frac{1}{M}\right) &{\tilde s}^\prime,\; r^\prime\; {\rm fixed}\\
    O\left(\frac{1}{M}\right) &s^\prime,\; {\tilde r}^\prime\; {\rm fixed}
    \end{cases}\\
D^2_{sr}&\approx&\begin{cases}-{\displaystyle{i^K\sqrt{1-x}}}
  \frac{\displaystyle{\exp\{2\pi i
      {\tilde s}^\prime x\}-\exp\{2\pi i
      ({\tilde s}^\prime-{\tilde t}^\prime)}\}}{
    \displaystyle{2\pi i(
      {\tilde s}^\prime (1-x)-{\tilde t}^\prime)}},
  & {\tilde s}^\prime,\; {\tilde t}^\prime\; {\rm fixed}\\
 - {\displaystyle{(-i)^K\sqrt{1-x}}}
  \frac{\displaystyle{\exp\{2\pi is^\prime x\}
      -\exp\{2\pi i(s^\prime-t^\prime}\}}{
    \displaystyle{2\pi i(
      s^\prime (1-x)-t^\prime)}},
  & s^\prime,\; t^\prime\; {\rm fixed}\\
  O\left(\frac{1}{M}\right) &{\tilde s}^\prime,\; t^\prime; {\rm fixed}\\
    O\left(\frac{1}{M}\right) &s^\prime,\; {\tilde t}^\prime\; {\rm fixed}
  \end{cases}
  \eea
  Looking back at Eq (\ref{ffoverlap}), we see that the overlap in a free
  fermion  system is the exponential of a bilinear form in the coherent state
  parameters. with coefficients built from the matrices $D^1,D^2$. The
  simplification  that the matrix elements, linking modes near $3M/4$
  to modes near $M/4$, vanish in the string limit means that there are no
  corresponding terms in the bilinear form. The absence of these terms
  means that the vertex conserves ${\hat P}$ as well as $Q$. Thus in the
  bosonized formulation the zero modes $a_0$ and ${\tilde a}_0$ are
  also  separately conserved.  We can call the modes near $3M/4$
  right moving (${\hat P}>0$ and those near $M/4$ left moving (${\hat P}<0$).
  The decoupling
  also implies that the vertex factorizes into a factor involving only right
  moving modes times a factor involving only left moving modes.

As we have discussed each of the labels $s^\prime, r^\prime, t^\prime$,
  as well as the tilde labels, are integers plus a fractional part $\alpha$
  (${\tilde\alpha}$) which
  depends on the charge $Q$ of the corresponding chain:
  \bea
  \alpha&=&\begin{cases} \frac{1}{2}& Q\in 4\mathbb{Z}\\
    0& Q\in 2+ 4\mathbb{Z}\\
   -\frac{1}{4} & Q\in 1 +  4\mathbb{Z}\\
    \frac{1}{4} & Q\in -1 +  4\mathbb{Z}
  \end{cases},\qquad
  {\tilde \alpha} = \begin{cases} \frac{1}{2}& Q\in 4\mathbb{Z}\\
    0& Q\in 2+ 4\mathbb{Z}\\
   \frac{1}{4} & Q\in 1 +  4\mathbb{Z}\\
   - \frac{1}{4} & Q\in -1 +  4\mathbb{Z}
  \end{cases}
  \eea
  Note that the fractional parts of the right and left moving modes
  are the same when $Q$ is even, and they are opposites when $Q$ is odd.
  
  Conservation of $Q$ implies selection rules for these fractional modes.
  Once the fractional modes are assigned to two of the chains, the third
  is fixed. The following transitions are allowed:
  \bea
&&  (1/2,1/2)\to 1/2,\qquad  (1/2,0)\to 0,\qquad  (1/2,1/4)\to 1/4,\qquad
  (1/2,-1/4)\to -1/4,\nonumber\\
&&  (0,0)\to 1/2,\qquad  (0,1/4)\to -1/4, \qquad    (0,-1/4)\to 1/4,\nonumber\\
&& (1/4,1/4)\to 0,\qquad (1/4,-1/4)\to 1/2,\qquad  (-1/4,-1/4)\to 0.
  \eea
where the fraction refers to the fractional part of the right moving modes.
  The selection rules involving only integer and half integer modes match those
  of the RNS spinning string model, with half odd integers
  the NS sector and integers the R sector. In fact,
  the overlap calculations in these sectors
  also coincide with those in the RNS formalism.
  The presence of quarter integer modes is a peculiar
 feature of the Heisenberg spin chain not found in the RNS string. 

  It is worth noting that these selection rules introduce a complication
  in manipulations with $D^1,D^2$, required to evaluate the full vertex,
  compared to some of the forbidden
  transitions. First notice that $r^\prime$ in the numerator of $D^1$
  and $t^\prime$ in the numerator of $D^2$ can be replaced by their
  fractional parts. Then, for example,
  the forbidden transition $(0,0)\to 0$ would allow an overall
  factor of $1-e^{2\pi i s^\prime x}$ to be removed from both $D^1$ and
  $D^2$. But for the allowed transitions, e.g. $(1/2,1/2)\to 1/2$ or
  $(0,0)\to 1/2$, one of these factors is
  $1-e^{2\pi i s^\prime x}$ but the other is $1+e^{2\pi i s^\prime x}$,
  so this simplification is not possible.

  To flesh this out a little, specialize to the case where the fractional
  parts of all modes are $1/2$. Then the matrix $D$ has elements
\bea
D^1_{sr}&\approx& \begin{cases}-{\displaystyle{\sqrt{x}}}
  \frac{\displaystyle{
      1+e^{2\pi i{\tilde s}^\prime x}}}{
    \displaystyle{2\pi i(
      {\tilde s}^\prime x-{\tilde r}^\prime )}},
  & {\tilde s}^\prime,\; {\tilde r}^\prime\; {\rm fixed}\\
  -{\displaystyle{\sqrt{x}}}
  \frac{\displaystyle{1+e^{2\pi is^\prime x}}}{
    \displaystyle{2\pi i(
      s^\prime x-r^\prime)}},
  & s^\prime,\; r^\prime\; {\rm fixed}\\
  O\left(\frac{1}{M}\right) &{\tilde s}^\prime,\; r^\prime\; {\rm fixed}\\
    O\left(\frac{1}{M}\right) &s^\prime,\; {\tilde r}^\prime\; {\rm fixed}
    \end{cases}\\
D^2_{sr}&\approx&\begin{cases}-{\displaystyle{i^K\sqrt{1-x}}}
  \frac{\displaystyle{e^{2\pi i
      {\tilde s}^\prime x}-1}}{
    \displaystyle{2\pi i(
      {\tilde s}^\prime (1-x)-{\tilde t}^\prime)}},
  & {\tilde s}^\prime,\; {\tilde t}^\prime\; {\rm fixed}\\
 - {\displaystyle{(-i)^K\sqrt{1-x}}}
  \frac{\displaystyle{e^{2\pi is^\prime x}
      -1}}{
    \displaystyle{2\pi i(
      s^\prime (1-x)-t^\prime)}},
  & s^\prime,\; t^\prime\; {\rm fixed}\\
  O\left(\frac{1}{M}\right) &{\tilde s}^\prime,\; t^\prime; {\rm fixed}\\
    O\left(\frac{1}{M}\right) &s^\prime,\; {\tilde t}^\prime\; {\rm fixed}
  \end{cases}
  \eea
  Had $s^\prime$ and ${\tilde s}^\prime$ been integers instead of half odd
  integers the matrix $(D^1|D^2)$ could be written as a diagonal matrix
  with entries $(1+\exp\{2\pi i{\tilde s}^\prime x\})$ times a matrix with
elements   $1/(s^\prime x-r^\prime)$ or $1/(s^\prime (1-x)-t^\prime)$,
which can be inverted by a method developed by J. Goldstone using analytic
functions, or by the Neumann function methods of Mandelstam. When on shell
methods were applied to the Neveu-Schwarz vertex by Hornfeck \cite{hornfeck},
he obtained
a double sum over the simple closed form expressions for the bosonic string
or Green-Schwarz three vertex. It is not mandatory that the answer be in
closed form, but it is mandatory that the matrix $(D^1|D^2)$ reduce, in the
  limit $M\to\infty$ with $x=K/M$ fixed, to the overlap matrices used 
  to derive the string scattering amplitudes, and they do \cite{berkovits}.

  On a more hopeful note, switching to bosonized language makes the overlap
  calculation treatable by Mandelstam's methods. And this can easily be
  extended to the conjectured $\Delta\neq0$ case. The conjecture, supported by
  several explicit calculations, is that the only change for $\Delta\neq0$
  is a change in the radius of compactification
  \bea
  R^2_\Delta=\frac{1}{2\pi T_0}\frac{\pi}{\mu}=\alpha^\prime\frac{\pi}{\mu},
  \qquad \Delta=-\cos\mu
  \eea
  \cite{spacebits,gilesmt}. We see that the decompactification limit
  is $\Delta\to -1$. In this limit the chains with odd $M$ (odd $Q$)
  acquire infinite energy in the $R\to\infty$ limit and hence will
  decouple from finite energy processes. The conjectured equivalence
  to a compactified boson is
  very plausible in a neighborhood of $\Delta=0$. If it holds for all $\Delta$
  in the critical range $-1<\Delta<1$, the scattering amplitudes of closed
  string theory will apply, including graviton graviton scattering in the
  decompactification limit.
\vskip12pt
\noindent  {\bf Acknowledgements}: I thank Lars Brink and Nathan Berkovits for helpful
  comments, and Sourav Raha for illuminating discussions. 
  This work was supported in part by
  the Department of Energy under Grant No.DE-SC0010296.
%%%%%%%%%


\begin{thebibliography}{1}
\bibitem{spacebits}
  C.~B.~Thorn,
  ``Space from String Bits,''
  JHEP {\bf 1411} (2014) 110
  [arXiv:1407.8144 [hep-th]].
  %%CITATION = ARXIV:1407.8144;%%
\bibitem{gilest}
  R.~Giles and C.~B.~Thorn,
  ``A Lattice Approach To String Theory,''
  Phys.\ Rev.\  D {\bf 16} (1977) 366.
  %%CITATION = PHRVA,D16,366;%%
\bibitem{thornsakh}
  C.~B.~Thorn, ``Reformulating string theory with the 1/N expansion,'' 
[arXiv: hep-th/9405069]; see also 
%invited talk given at The First International A. D. Sakharov
 % Conference on Physics, Moscow, 1991;
  %``Reformulating string theory with the 1/N expansion,''
  {\it Sakharov memorial lectures in physics, vol. 1}, Edited by L. V. Keldysh 
and V. Ya. Fainberg, (Nova Science Publishers, Commack, NY, 1992) 447-453.
%and Florida Univ. Gainesville - UFIFT-HEP-91-23 (91,rec.Dec.) 9 p. (200590)
 % \cite{Gliozzi:1976qd}
\bibitem{gso}
  F.~Gliozzi, J.~Scherk and D.~I.~Olive,
  ``Supersymmetry, Supergravity Theories and the Dual Spinor Model,''
  Nucl.\ Phys.\ B {\bf 122} (1977) 253; P.~Ramond,
  ``Dual Theory for Free Fermions,''
  Phys.\ Rev.\  D {\bf 3} (1971) 2415;  A.~Neveu and J.~H.~Schwarz,  
``Factorizable dual model of pions,''  
Nucl.\ Phys.\  B {\bf 31} (1971) 86; A.~Neveu, J.~H.~Schwarz and C.~B.~Thorn,
  ``Reformulation of the Dual Pion Model,''
  Phys.\ Lett.\  B {\bf 35} (1971) 529.
C.~B.~Thorn,
  ``Embryonic Dual Model for Pions and Fermions,''
  Phys.\ Rev.\  D {\bf 4} (1971) 1112;
A.~Neveu and J.~H.~Schwarz,  ``Quark Model of Dual Pions,''
 Phys.\ Rev.\  D {\bf 4} (1971) 1109.
  %%CITATION = NUPHA,B122,253;%%
  %966 citations counted in INSPIRE as of 14 Feb 2014
\bibitem{thooftlargen}
G. 't Hooft, {\sl Nucl. Phys.} {\bf B72} (1974) 461.
\bibitem{sunthorn}
  S.~Sun and C.~B.~Thorn,
  ``Stable String Bit Models,''
  Phys.\ Rev.\ D {\bf 89} (2014) 105002
  [arXiv:1402.7362 [hep-th]].
  %%CITATION = ARXIV:1402.7362;%%
\bibitem{chensun}
  G.~Chen and S.~Sun,
  ``Numerical Study of the Simplest String Bit Model,''
  Phys.\ Rev.\ D {\bf 93} (2016) %no.10,
106004
  %doi:10.1103/PhysRevD.93.106004
  [arXiv:1602.02166 [hep-th]].
  %% CITATION = doi:10.1103/PhysRevD.93.106004;%%
\bibitem{songge}
Songge Sun,
``Aspects of Stable String Bit Models'',
Ph.D. thesis, University of Florida, 2019.
 \bibitem{goddardgrt}
  P.~Goddard, C.~Rebbi, C.~B.~Thorn,
  ``Lorentz covariance and the physical states in dual resonance models,''
  Nuovo Cim.\  {\bf A12 } (1972)  425-441.
%\bibitem{goddardgrt}
P.~Goddard, J.~Goldstone, C.~Rebbi and C.~B.~Thorn,
``Quantum dynamics of a massless relativistic string,''
 Nucl.\ Phys.\  B {\bf 56} (1973) 109.
  %%CITATION = NUPHA,B56,109;%%
\bibitem{bergmantsubit}
  O.~Bergman and C.~B.~Thorn,
  ``String bit models for superstring,''
  Phys.\ Rev.\ D {\bf 52} (1995) 5980
  [hep-th/9506125];
  %%CITATION = HEP-TH/9506125;%%
  %42 citations counted in INSPIRE as of 15 Oct 2013
%\cite{Bardakci:1970nb}
\bibitem{bardakcih}
  K.~Bardakci and M.~B.~Halpern,
  ``New dual quark models,''
  Phys.\ Rev.\ D {\bf 3} (1971) 2493;
  %%CITATION = PHRVA,D3,2493;%%
  %301 citations counted in INSPIRE as of 14 Feb 2014
%\bibitem{greenschwarz}
  M.~B.~Green and J.~H.~Schwarz,
  ``Supersymmetrical Dual String Theory,''
  Nucl.\ Phys.\ B {\bf 181} (1981) 502.
  %%CITATION = NUPHA,B181,502;%%
  %322 citations counted in INSPIRE as of 14 Feb 2014
\bibitem{thornprotobits}
  C.~B.~Thorn,
  ``1/N Perturbations in Superstring Bit Models,''
  Phys.\ Rev.\ D {\bf 93} (2016) no.6,  066003
 % doi:10.1103/PhysRevD.93.066003
  [arXiv:1512.08439 [hep-th]];
  %%CITATION = doi:10.1103/PhysRevD.93.066003;%%
 %\cite{Thorn:2016syo}
%\bibitem{Thorn:2016syo}
C.~B.~Thorn,
``Protostring Scattering Amplitudes,''
Phys. Rev. D \textbf{94} (2016) no.10, 106009
doi:10.1103/PhysRevD.94.106009
[arXiv:1607.04237 [hep-th]].
%0 citations counted in INSPIRE as of 18 Sep 2020
\bibitem{bethe}
H. A. Bethe, {\sl Z. Phys.} {\bf 61} (1930) 206.
\bibitem{yangyang}
C. N. Yang and C. P. Yang, {\sl Phys. Rev.} {\bf 150} (1966) 321; {\sl Phys.
  Rev.} {\bf 150} (1966) 327.
%\cite{Lieb:1961fr}
\bibitem{liebsm}
E.~H.~Lieb, T.~Schultz and D.~Mattis,
``Two soluble models of an antiferromagnetic chain,''
Annals Phys. \textbf{16} (1961), 407-466
doi:10.1016/0003-4916(61)90115-4
%636 citations counted in INSPIRE as of 06 Sep 2020
% \cite{gilesmt}
\bibitem{gilesmt} 
  R.~Giles, L.~D.~McLerran and C.~B.~Thorn,
  ``The String Representation For A Field Theory With Internal Symmetry,''
  Phys.\ Rev.\ D {\bf 17}, 2058 (1978).
  %%CITATION = PHRVA,D17,2058;%%
  %26 citations counted in INSPIRE as of 12 Feb 2014
\bibitem{brinkgs}
  M.~B.~Green, J.~H.~Schwarz and L.~Brink,
  ``Superfield Theory of Type II Superstrings,''
  Nucl.\ Phys.\ B {\bf 219} (1983) 437.
  %%CITATION = NUPHA,B219,437;%%
  %178 citations counted in INSPIRE as of 22 Jul 2014%\cite{Thorn:2009qp}
%\cite{Mandelstam:1974hk}
\bibitem{mandelstamlc}  
S.~Mandelstam,
 ``Interacting String Picture of Dual Resonance Models,''
  Nucl.\ Phys.\  B {\bf 64} (1973) 205;
 %%CITATION = NUPHA,B64,205;%%
%\bibitem{mandelstamnsr}
  %S.~Mandelstam,
  ``Interacting String Picture of the Neveu-Schwarz-Ramond Model,''
 Nucl.\ Phys.\  B {\bf 69} (1974) 77.
 %%CITATION = NUPHA,B69,77;%%
%\cite{Mandelstam:1974fb}
%\bibitem{Mandelstam:1974fb}
%S.~Mandelstam,
``Lorentz Properties of the Three-String Vertex,''
Nucl. Phys. B \textbf{83} (1974), 413-439
doi:10.1016/0550-3213(74)90266-1
% 112 citations counted in INSPIRE as of 07 Sep 2020
% \bibitem{thornps}
% C.~B.~Thorn, ``Reformulating string theory with the 1/N expansion,''
%postscript, [arXiv: hep-th/9405069]
  %52 citations counted in INSPIRE as of 15 Oct 2013
  %\cite{thornsubstructure}
 %\cite{Hornfeck:1987wt}
\bibitem{hornfeck}
K.~Hornfeck,
``Three Reggeon Light Cone Vertex of the Neveu-schwarz String,''
Nucl. Phys. B \textbf{293} (1987), 189
doi:10.1016/0550-3213(87)90068-X
% 15 citations counted in INSPIRE as of 07 Sep 2020
%\cite{Berkovits:1987gp}
\bibitem{berkovits}
N.~Berkovits,
``Supersheet Functional Integration and the Interacting Neveu-schwarz String,''
Nucl. Phys. B \textbf{304} (1988), 537-556
doi:10.1016/0550-3213(88)90642-6
% 37 citations counted in INSPIRE as of 07 Sep 2020
\end{thebibliography}
\end{document}